%% file: Manuscript.tex
\RequirePackage{fix-cm}
\documentclass[smallcondensed]{svjour3}     
\smartqed  
\usepackage{graphicx}
\usepackage[printonlyused]{acronym}
\usepackage{caption}		
\usepackage[table]{xcolor}
\definecolor{lightgray}{gray}{0.9}
\newsavebox\mybox
\usepackage{subfigure}
\usepackage{amsmath}
\usepackage{multirow}
\usepackage[ruled]{algorithm2e} 

\usepackage[toc,page]{appendix}

\usepackage{xcolor}
\usepackage{listings}
\usepackage{comment}
\lstdefinestyle{appendixcode}{
  basicstyle=\ttfamily\footnotesize,
  columns=fullflexible,
  breaklines=true,
  breakatwhitespace=true,
  keepspaces=true,
  showstringspaces=false,
  frame=single,
  rulecolor=\color{black!20},
  xleftmargin=0.5em,
  xrightmargin=0.5em,
  aboveskip=0.8em,
  belowskip=0.8em
}

\lstdefinelanguage{ttl}{
  morekeywords={@prefix,a},
  sensitive=true,
  morecomment=[l]{\#},
  morestring=[b]"
}

\begin{document}

\title{Publication and Maintenance of Relational Data in Enterprise Knowledge Graphs (Revised Version)
}

\titlerunning{Publication and Maintenance of Relational Data in Enterprise Knowledge Graphs (Revised Version)}        

\author{V\^ania Maria Ponte Vidal         \and
        Val\'eria Magalh\~aes Pequeno \and
         Marco Antonio Casanova \and
         Narciso Arruda \and
         Carlos Brito 
}

\authorrunning{V.M.P. et al} 

\institute{V\^ania Maria Ponte Vidal, Narciso Arruda, Carlos Brito \at
              Departamento de Computa\c{c}\~ao, Universidade Federal do Cear\'a,  Av. da Universidade, 2853, Fortaleza, Cear\'a,  60020-181, Brazil\\
              \email{\{vvidal, narciso, carlos\}@lia.ufc.br}       
           \and
           Val\'eria Magalh\~aes Pequeno \at
           TechLab, Departamento de Ci\^{e}ncias e Tecnologias, Universidade Aut\'{o}noma de Lisboa Lu\'{i}s de Cam\~{o}es,  Rua Santa Marta, Palácio Dos Condes Do Redondo 56, 1169-023, Lisboa, Portugal \\
            \email{vpequeno@autonoma.pt}   
           \and
              Marco Antonio Casanova \at
              Departmento de Informatica, 
              Pontifícia Universidade Católica do Rio de Janeiro, 
              R. Marquês de São Vicente, 225, Rio de Janeiro, 22451-045, Brazil\\
              \email{casanova@inf.puc-rio.br}
}

\date{Received: date / Accepted: date}

\maketitle

\begin{abstract}
Enterprise knowledge graphs (EKGa) are a novel paradigm for consolidating and semantically integrating large numbers of heterogeneous data sources into a comprehensive dataspace. The main goal of an EKG is to provide a data layer that is semantically connected to enterprise data, so that applications can have integrated access to enterprise data sources through that semantic layer. To make legacy relational data sources accessible through the organization's knowledge graph, it is necessary to create an RDF view of the underlying relational data (RDB2RDF view). An RDB2RDF view can be materialized to improve query performance and data availability. However, a materialized RDB2RDF view must be continuously maintained to reflect updates over the relational database. 
This article proposes a formal framework for the construction of the materialized data graph for an RDB2RDF view, and for 
incremental maintenance of the view's data graph. 
The article also presents an architecture and algorithms for implementing the proposed framework. 
\keywords{View Maintenance \and RDF view \and Linked Data \and Relational Database \and RDB2RDF mapping \and enterprise knowledge graph}
\end{abstract}

\newcommand{\Sii}{\ensuremath{\textbf{\textit{S}}}}		
\newcommand{\Rn}{\ensuremath{\textit{R}}}			
\newcommand{\At}{\ensuremath{\textit{A}}}             		
\newcommand{\Ri}{\ensuremath{\Omega}}			
\newcommand{\Fk}{\ensuremath{\textit{F}}}             		
\newcommand{\Rc}{\ensuremath{\textit{\textbf{R}}}}		
\newcommand{\St}{\ensuremath{\sigma}}				


\newcommand{\Oi}{\ensuremath{\textit{\textbf{O}}}}		
\newcommand{\Cn}{\ensuremath{\textit{C}}}               	
\newcommand{\Pt}{\ensuremath{\textit{P}}}             		
\newcommand{\Ov}{\ensuremath{\textit{O}_V}}			
\newcommand{\Ci}{\ensuremath{\Sigma}}				

\newcommand{\Vi}{\ensuremath{\mathcal{W}}}				

\newcommand{\Ti}{\ensuremath{\textit{r}}}				
\newcommand{\uri}{\ensuremath{\textit{s}}}		         
\newcommand{\Ts}{\ensuremath{\textit{\textbf{T}}}}		
\newcommand{\Qd}{\ensuremath{\textit{Q}}}			

\newcommand{\Ru}{\ensuremath{\Psi}}			         
\newcommand{\Mv}{\ensuremath{\textit{M}}}			
\newcommand{\Ai}{\ensuremath{\textit{\textbf{A}}}}		
\newcommand{\Qn}{\ensuremath{\textit{Q}}}			
\newcommand{\Do}{\ensuremath{\textit{D}}}                 	
\newcommand{\vl}{\ensuremath{\textit{v}}}                 		

\newcommand{\sel}{\ensuremath{\delta}}            		

\newcommand{\Ca}{\ensuremath{\phi}}                 		

\newcommand{\tm}{\ensuremath{\textit{t}}}                 	

\newcommand{\up}{\ensuremath{\textit{u}}}                 	
\newcommand{\Dm}{\ensuremath{\Delta^{-}}}                 	
\newcommand{\Dp}{\ensuremath{\Delta^{+}}}                 	

\newcommand{\bs}[1]{\boldsymbol{#1}}
\newcommand{\Pc}{\ensuremath{\mathcal{P}}}                 		
\newcommand{\Lc}{\ensuremath{\mathcal{L}}}                 		
\newcommand{\Pcb}{\ensuremath{\bs{\mathcal{P}}}}
\newcommand{\Sc}{\ensuremath{\mathcal{S}}}                 		


\acrodef{CCA}[CCA]{Class Correspondence Assertion}
\acrodef{DCA}[DCA]{Datatype Property Correspondence Assertion}
\acrodef{OCA}[OCA]{Object Property Correspondence Assertion}
\acrodef{LD}[LD]{Linked Data}
\acrodef{LOD}[LOD]{Linked Open Data}
\acrodef{RDF}[RDF]{Resource Description Framework}
\acrodef{LDA}[LDA]{Linked Data Application}
\acrodef{CA}[CA]{Correspondence Assertion}
\acrodef{TR}[TR]{Transformation Rule}
\acrodef{UUID}[UUID]{Universally Unique IDentifier}
\acrodef{OBDA}[OBDA]{Ontology-Based Data Access}


\section{Introduction}
\label{s-int}

Enterprise knowledge graphs (EKGs) are a novel paradigm for consolidating and semantically integrating large numbers of heterogeneous data sources into a comprehensive dataspace \cite{Pan18}. 
The main goal of an EKG is to provide a unified data layer that is semantically connected to enterprise data, enabling applications to access data sources through the semantic layer. 
In this way, an EKG can support unplanned ad hoc queries and data exploration without requiring complex, time-consuming data preprocessing. 

A key element of an EKG is the ontology, which describes all the information present within the knowledge graph. In an EKG, the ontology serves as a semantic layer that combines and enriches information from data sources into a unified view. Therefore, users and applications can query the ontology transparently, without having to deal with complex data source schemes.

In many large organizations, the information in a knowledge graph is derived from legacy relational databases. To make this information accessible through the organization's knowledge graph, it is necessary to create an RDF view on top of the underlying relational data, called an RDB2RDF view. 
This view is specified by a set of mappings that translate source data into the organization's ontology vocabulary~\cite{JODS-2008}.  

The content of an RDB2RDF view can be materialized to improve query performance and data availability. In the \textit{materialization} of the data graph for an RDB2RDF view  \Vi, a set of mappings \Mv\ is used to translate a state of the input relational source \Sii\ into the corresponding view state, which is a data graph \Ts.
Then, the answer to a query $Q$ over the RDB2RDF view is computed by executing $Q$ over \Ts. 

However, to be useful, a materialized RDB2RDF view must be continuously maintained to reflect updates to the source.
Basically, there are two strategies for materialized view maintenance. 
\textit{Rematerialization} recomputes view data at pre-established times, whereas \textit{incremental maintenance} periodically modifies part of the view data to reflect updates to the database. 
It has been shown that incremental maintenance generally outperforms full view rematerialization~\cite{Abiteboul98b,AFP00,CW91}. 
Another advantage of incremental maintenance is that it enables \textit{live synchronization} of the view with the data source, as the view is up-to-date with only a small delay. 
This is an important property when the data source is frequently updated.

A popular strategy used by large \ac{LOD} datasets to maintain materialized RDF views is to compute and publish changesets that capture the differences between two states of the dataset. 
Applications can then download the changesets and synchronize their local replicas. 
For instance, DBpedia~\cite{Dbpedia} and LinkedGeoData~\cite{LG} publish their changesets in a public folder.
A strategy based on changesets, computed for each update over the relational data source, can also be devised to maintain a materialized RDB2RDF view incrementally. 

The diagram in Figure~\ref{f-mv} describes the problem of computing a correct changeset for a materialized RDB2RDF view \Vi, when an update \up\ occurs in the source relational database \Sii. 
In the diagram of Figure~\ref{f-mv}, assume that:  

\begin{figure}
	\centering
	\includegraphics[width=0.5\textwidth]{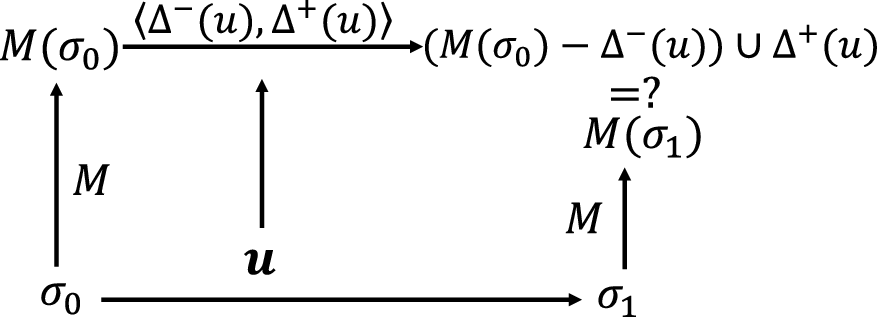}
	\caption{The problem of computing the correct changeset for RDB2RDF View.}
	\label{f-mv}
\end{figure}

\begin{enumerate}
	\item \Mv\ is the set of mappings that materialize view \Vi; 
	\item $\sigma_0$ and $\sigma_1$ are the states of \Sii\ respectively before and after the update \up; and 
	\item \Mv($\sigma_0$) and \Mv($\sigma_1$) are the materializations of \Vi\ respectively at $\sigma_0$ and $\sigma_1$. 
\end{enumerate}

Let \up\ be an update and $\sigma_0$ and $\sigma_1$ be the database states before and after \up, respectively.
A \textit{correct changeset} for \Vi\ w.r.t. \up, $\sigma_0$ and $\sigma_1$
is a pair $\langle$\Dm(\up), \Dp(\up)$\rangle$, 
where \Dm(\up) is a set of triples removed from \Vi\ and 
\Dp(\up) is a set of triples added to \Vi, 
that satisfies the following restriction (see Figure~\ref{f-mv}):

\begin{equation} \label{eq:mv1}
             \Mv(\sigma_1) = (\Mv(\sigma_0) - \Dm(\up)) \cup \Dp(\up)
\end{equation}

\noindent Putting this in words, the changeset $\langle$\Dm(\up), \Dp(\up)$\rangle$ is correctly computed  
iff the new view state $(\Mv(\sigma_0) - \Dm(\up)) \cup \Dp(\up)$,  
computed with the help of the changeset, and the new view state $\Mv(\sigma_1)$, 
obtained by rematerializing the view using the view mappings are identical. 

The computation of the changesets depends on the update \up, 
the initial and final database states, $\sigma_0$ and $\sigma_1$,
and the view mappings \Mv.
However, the notation for changesets indicates only the update \up, to avoid a cumbersome notation,
since we may consider the database states as the context for \up,
and the view mappings as fixed for the database in question.

In this paper, we propose a solution for computing changesets for RDB2RDF views, focusing on three challenges: (1) Simplicity, that is, we want to minimize the complexity associated with the creation of the infrastructure responsible for the construction of the changesets; (2) Efficiency of operation, that is, we want to identify the minimal data that permits the construction of changesets that correctly maintain the view in the face of updates on the source database; (3) Self-maintenance of RDB2RDF views, that is, we want to compute the changeset based solely on the update and the source database state. 

Our solution is based on three key ideas. 
First, we note that RDB2RDF views typically have the so-called \textit{object-preserving property}. 
That is, they preserve the base entities (objects) of the source database, rather than creating new entities from the existing ones~\cite{Motschnig00}. 
Therefore, an instance of the RDF view corresponds to a database tuple, and both represent the same real-world object. 
By restricting ourselves to this class of views, we can precisely identify the tuples relevant to a data source update w.r.t. an RDB2RDF view. 
Those are the tuples whose corresponding RDF instances (i.e., their RDF-image) might have been affected by the update. 
In the proposed strategy, only the portion of the RDF image associated with the relevant tuples should be rematerialized. 
Therefore, efficient maintenance is possible for object-preserving views.  

Second, we introduce a formalism 
that can specify object-preserving view mappings. 
The formalism makes it easier to understand the semantics of the mapping, 
and it provides sufficient information to support: (1) The identification of the tuples that are relevant to a data source update; (2) The automatic construction of the procedures that actually compute the changesets; (3) A rigorous proof of the correctness of the approach.

Third, in the proposed framework, the content of an RDB2RD view is stored in an RDF dataset containing a set of named graphs that describe the context in which the triples were produced. The main reason for separating triples into distinct named graphs is that duplicated triples, produced by tuples in different relations, will be in different named graphs. 
This is an important property for proving that the proposed framework correctly computes the changesets.

In~\cite{Vania2017}, the authors proposed a framework 
based on triggers that provides live synchronization of RDB2RDF views. 
The RDB2RDF view synchronization tool keeps the RDB2RDF view in sync with the relational database.
In that framework, an RDB2RDF view is specified by a set of \acp{CA} and a set of triggers is responsible for computing and publishing the changesets.
The proposed strategy first identifies the relations in the source database relevant to the RDB2RDF view, that is, those whose updates might affect the RDB2RDF view's state. 
For each type of update operation (insert, delete, and update) on a relevant relation, two triggers are defined:(1) BEFORE Trigger: fired immediately before the update to compute \Dm; (2) AFTER Trigger: fired immediately after the update to compute \Dp. 

This article extends the work reported in~\cite{Vania2017} in two directions.
First, it proposes a formalism for specifying Object-Preserving RDB2RDF Views. 
The formalism used to represent transformation rules is based on first-order logic and has been widely adopted in \ac{OBDA}~\cite{SAM14}, data exchange~\cite{MLBA14}, and data integration~\cite{Lenzerini02} systems. 
Second, it proposes a formal framework for computing the correct changeset for an RDB2RDF view specified by a set of transformation rules. 
In the proposed framework, a changeset is computed solely from the source update and the source state before and after the update, and hence no access to the materialized view is required. 
This is important when the view is maintained externally~\cite{VSM05}, because accessing a remote data source may be too slow.

The remainder of this article is organized as follows. 
Section~\ref{s-rw} discusses related work. 
Section~\ref{s-map} presents the formalism used for specification of object-preserving RDB2RDF views.
Section~\ref{s-cs} introduces the case study that is used throughout the article. 
Section~\ref{s-mgr} formalizes the materialization of the data graph for an RDB2RDF view.
Section~\ref{s-fw} presents a formal framework for computing the correct changeset for an RDB2RDF view. 
Section~\ref{s-conc} presents the conclusions. 
\section{Related Work}
\label{s-rw}

The Incremental View Maintenance problem has been extensively studied in the literature for relational views~\cite{CW91,Blakeley86}, object-oriented views~\cite{AFP00,AFP03}, semi-structured views~\cite{LD00,Zhao17}, and XML Views~\cite{VLAC08,JL10,Fegaras2011}. 

Most work on relational view maintenance proposes an algorithm that computes the changes to the materialized view when the base relations are updated. 
The work in~\cite{CW91} is the closest related to ours. It shows that using triggers is effective for incremental maintenance because most of the work is done at view definition time. However, the method the authors propose does not support efficient maintenance of views with duplicates.

Blakeley et al.~\cite{Blakeley86} propose two solutions for the problem of duplicates in relational views. 
In the first solution, an extra column is added to the view table to count occurrences of each view tuple. 
This approach is not entirely suitable, as the trigger-generation process may become quite complex. 
The idea is to have a tool that automatically generates triggers, as in~\cite{CW91}.
The second solution proposed in~\cite{Blakeley86} ensures that a view does not contain duplicates by requiring that each base table have a key in the view. 
Oracle implements an incremental refresh mechanism using this approach. 
However, we could not apply this artifice to RDB2RDF views, in a straightforward matter, since the RDF view is obviously a set of triples. 

The main idea that distinguishes our approach from previous work on relational view maintenance is to explore the object-preservation property of typical RDB2RDF views, which enables the identification of the specific tuples relevant to a data source update. 
Only the portion of the view associated with the relevant tuples should be re-materialized. 
In particular, we may characterize the approach as ``tracking the relevant tuples in the pivot relations for a given update" rather than ``tracking the updated triples in the view for a given update," as is done in relational view maintenance.  

Konstantinou et al.~\cite{KSK15} investigate the problem of incremental generation and storage of an RDF graph that is the result of exporting relational database contents.
Their strategy, which we call \textit{partial re-materialization}, requires annotating each triple with the mapping definition that generated it. 
In this case, when one of the source tuples changes (i.e., a table appears to be modified), the triples map definition will be executed for all tuples generated using the affected table and, thus, all triples generated using the affected tables are rematerialized. 
By contrast, in our approach, we can identify which tuples in the affected tables are affected by the update, and only those are rematerialized.  

Endris et al.~\cite{Endris2015} introduce an approach for interest-based RDF update propagation that consistently maintains a full or partial replication of large \ac{LOD} datasets. 
Faisal et al.~\cite{Faisal16} present an approach for dealing with co-evolution, which refers to the mutual propagation of the changes between a replica and its origin dataset. 
Both approaches rely on the assumption that either the source dataset provides a tool to compute a changeset in real-time or a third-party tool can be used for this purpose. 
Therefore, the contribution of this article is complementary and relevant to satisfy their assumption.  
The works in~\cite{Papavasileiou13,Roussakis15,ZTC11} address the problem of change detection between versions of \ac{LOD} datasets.
In~\cite{ZTC11}, a low-level change detection approach is used to report simple insertion/deletion operations. 
In~\cite{Papavasileiou13,Roussakis15}, a high-level change detection approach is used to provide deltas that are more readable to humans. 
Despite their contributions to understanding and analysing the dynamics of Web datasets, these techniques cannot be applied to compute changesets for RDB2RDF views.

Vidal et al.~\cite{VCC13} proposed an incremental maintenance strategy, based on triggers, for RDF views defined on top of relational data. 
Their approach was further modified and enhanced to compute changesets for RDB2RDF views~\cite{Vania2017}.  
To the best of our knowledge,~\cite{Vania2017} was the first work that addressed the problem of computing changesets for RDB2RDF views.

This article extends the framework proposed in~\cite{Vania2017} to handle more complex mapping rules, represented by a formalism based on first-order logic and its semantics.  
Therefore, the formal framework proposed in Section~\ref{s-fw},
which differs considerably from that presented in~\cite{Vania2017}.

\section{Object Preserving RDB2RDF Views}\label{s-map}
\subsection{Basic Concepts and Notation}

As usual, a \textit{relation scheme} is denoted as \Rn[\At$_1$, $\dots$, \At$_n$].
The \textit{relational constraints} considered in this article consist of \textit{mandatory} (or not null) \textit{attributes, keys, primary keys} and \textit{foreign keys}. 
In particular, \Fk(\Rn:$L$, $S$:$K$) denotes a foreign key, named \Fk, that \textit{relate} \Rn\ and $S$, where $L$ and $K$ are lists of attributes from \Rn\ and $S$, respectively, with the same length. 

A \textit{relational schema} is a pair \Sii\ = (\Rc, \Ri), where \Rc\ is a set of relation schemes and \Ri\ is a set of relational constraints such that: (i) \Ri\ has a unique primary key for each relation scheme in \Rc; (ii) \Ri\ has a mandatory attribute constraint for each attribute which is part of a key or primary key; (iii) if \Ri\ has a foreign key of the form \Fk(\Rn:$L$, $S$:$K$), then \Ri\ also has a constraint indicating that $K$ is the primary key of $S$.
The \textit{vocabulary} of \Sii\ is the set of relation names, attribute names, and foreign key names used in \Sii.
Given a relation scheme \Rn[\At$_1$, $\dots$, \At$_n$] and a \textit{tuple variable} $t$ over \Rn, $t.\At_k$ denotes the \textit{projection} of $t$ over \At$_k$. 
\textit{Selections} over relation schemes are defined as usual.

Let \Sii\ = (\Rc,\Ri) be a relational schema and \Rn\ and $T$ be relation schemes of \Sii. 
A list \Ca\ = [\Fk$_1$, $\dots$, \Fk$_{n-1}$] of foreign key names of \Sii\ is a \textit{path from} \Rn\ \textit{to} $T$ iff there is a list \Rn$_1$, $\dots$, \Rn$_n$ of relation schemes of \Sii\ such that \Rn$_1$ = \Rn, \Rn$_n$ = $T$ and \Fk$_i$ \textit{relates} \Rn$_i$ \textit{and} \Rn$_{i+1}$. 
In this case, the tuples of \Rn\ \textit{reference tuples of} $T$ \textit{through} \Ca. 
A state \St\ of a relational schema \Sii\ assigns to each relation scheme \Rn\ of \Sii\ a relation \Rn(\St), in the usual way.
 
An \textit{ontology vocabulary}, or simply a \textit{vocabulary}, 
is a set of \textit{class names, object property names} 
and \textit{datatype property names}. 
An \textit{ontology} is a pair \Oi\ = ($V$,\Ci) such that 
$V$ is a vocabulary and \Ci\ is a finite set of formulae in $V$, 
the \textit{constraints} of \Oi. 
The constraints include the definition of the \textit{domain} and \textit{range} of a property, as well as \textit{cardinality constraints}, defined in the usual way.

\begin{definition}[Foreign Key]
Let $R$ and $S$ be relation schemes.
A foreign key is denoted by $F(R\!:\!L,\, S\!:\!K)$, where:
\begin{itemize}
  \item $F$ is the name of the foreign key;
  \item $L$ and $K$ are lists of attributes of $R$ and $S$, respectively;
  \item $|L| = |K|$.
\end{itemize}

\begin{enumerate}
\item The foreign key $F(R\!:\!L,\, S\!:\!K)$ relates the relation schemes $R$ and $S$;
that is, $R$ and $S$ are said to be \emph{related} by $F$.

\item The foreign key $F(R\!:\!L,\, S\!:\!K)$ induces a binary relationship between
tuples of $R$ and $S$ and may be traversed in either direction, from $R$ to $S$
or from $S$ to $R$.

\item A tuple $r \in R(\sigma)$ is related (or connected) to a tuple
$s \in S(\sigma)$ with respect to $F$ if and only if the values of the attributes
in $L$ of $r$ match the values of the attributes in $K$ of $s$.
\end{enumerate}
\end{definition}

\noindent\textbf{Note.}
Condition (iii) defines the satisfaction of a foreign key and is used to evaluate
whether two tuples are connected when traversing a path in a transformation rule,
independently of the traversal direction.

\begin{definition}[Relational Path]
Let $S = (R, \Omega)$ be a relational schema, and let 
$R$ and $T$ be relation schemes of $S$.

A list $\varphi = [F_1, \dots, F_{n-1}]$ of foreign keys of $S$
is a path from $R$ to $T$ if and only if there exists a list
$R_1, \dots, R_n$ of relation schemes of $S$ such that:

\begin{itemize}
    \item $R_1 = R$,
    \item $R_n = T$, and
    \item for each $i \in \{1, \dots, n-1\}$, the foreign key 
    $F_i$ relates $R_i$ and $R_{i+1}$.
\end{itemize}

\end{definition}
\begin{definition}[Relations of a Relational Path]
Let 
\[
\varphi = [F_1, \ldots, F_{n-1}]
\]
be a relational path, where each foreign key 
\[
F_i : R_i \rightarrow R_{i+1}
\]
connects relation schemes $R_i$ and $R_{i+1}$.

The operator $Relations(\cdot)$ applied to $\varphi$ returns 
the set of all relation schemes occurring along the path. Formally,

\[
Relations(\varphi) = \{ R_1, R_2, \ldots, R_n \}.
\]
\end{definition}

\begin{definition}[Connected to]
Given a database state $\sigma$, a tuple $r_1 \in R_1(\sigma)$ is
\emph{connected to} (or \emph{references}) a tuple $r_n \in R_n(\sigma)$
\emph{by following the path} $\phi$ if and only if there exists a sequence of
tuples $r_1,\ldots,r_n$ with $r_i \in R_i(\sigma)$ for $1 \leq i \leq n$ such that,
for each $i \in \{1,\ldots,n-1\}$, the tuples $r_i$ and $r_{i+1}$ are related with
respect to the foreign key $F_i$.
\end{definition}

\noindent\textbf{Note.}
Following a path $\phi$ allows traversal of each foreign key in either direction,
as specified in Definition~2

\begin{definition}[Related Tuples w.r.t.\ a Path]
Let $S=(\mathcal{R},\Omega)$ be a relational schema, and let $R^{*}$ and $R$
be relation schemes of $S$.
Let $\phi$ be a path relating $R^{*}$ and $R$.
Let $\sigma$ be a database state.
For a tuple $t \in R(\sigma)$, the set of tuples related to $t$ with respect
to path $\phi$, denoted by $P[\phi](t;\sigma)$, is defined as:
\[
P[\phi](t;\sigma)
=
\{\, p \in R^{*}(\sigma) \mid p \text{ is related to } t \text{ through } \phi
\text{ in } \sigma \,\}.
\]
\end{definition}

\noindent\textbf{Note.}
A tuple $p \in R^{*}(\sigma)$ is related to $t \in R(\sigma)$ through a path
$\phi$ if and only if there exists a sequence of tuples in $\sigma$ that
connects $p$ to $t$ by successively following each step of $\phi$, allowing
traversal of a foreign key or its inverse at each step.

\noindent\textbf{Note (Prefix of a Path).}
Let $\phi = [F_1,\ldots,F_n]$ be a path,
where each $F_i$ is a foreign key relating two relation schemes.
A \emph{prefix} of $\phi$ is any sequence
$\phi = [F_1,\ldots,F_k]$ with $0 \leq k \leq n$.
The relations connected by $\phi$ are the relation schemes reachable by successively following the steps in $\phi$,
independently of the direction of the foreign-key references.
%
%
\subsection{Specification of Object Preserving RDB2RDF View}

This section presents the formalism for specifying object-preserving RDB2RDF views. 
By restricting ourselves to this class of views, we can precisely identify the tuples relevant to a data source update w.r.t. an RDB2RDF view. 

Let \Oi\ = ($V$,\Ci) be a target ontology, that is, the organization's ontology, and let
\Sii\ = (\Rc,\Ri) be a relational schema, with vocabulary $U$. 
Let \textbf{\textit{X}} be a set of \textit{scalar variables} and \textit{\textbf{T}} be a set of tuple variables, disjoint from each other and from $V$ and $U$.

The formal definition of an RDB2RDF view is similar to that given in~\cite{SAM14,Poggi08}. 
An \textit{RDB2RDF view} is a triple \Vi = ($V$\!, \Sii, \Mv), where:
(i) $V$ is the vocabulary of the \textit{target ontology};
(ii) \Sii\ is the \textit{source relational schema}; and
(iii) \Mv\ is a set of \textit{mappings} between $V$ and \Sii, 
defined by transformation rules.

Intuitively, a view satisfies the object-preserving property
iff it preserves the base entities (objects) of the source database, 
rather than creating new entities from the existing ones~\cite{Motschnig00}. 
More precisely, a view \Vi = ($V$\!, \Sii, \Mv) satisfies 
the \textit{object-preserving property} iff:
\begin{itemize}
\item the instances of the classes in $V$ 
correspond to tuples in selected relations of \Sii,
which we call the \textit{pivot relations} of \Vi; 
\item the values of datatype properties in $V$ of these instances 
are given by (functions of) attributes in the corresponding tuples, 
or in related tuples;
\item the object properties in $V$ correspond to 
relationships between tuples in the pivot relations of the source database.
\end{itemize}

A \textit{transformation rule} of \Vi\ is an expression of the form \Cn($x$) $\leftarrow$ \Qn($x$) or of the form \Pt($x$,$y$) $\leftarrow$ \Qn($x$,$y$), where \Cn\ and \Pt\ are class and property names in $V$, and \Qn($x$) and \Qn($x$,$y$) are queries over \Sii\ 
whose target clauses contain one and two variables, respectively.

We adopt a formalism based on DATALOG~\cite{AHV95} for the specification of the queries \Qn($x$) and \Qn($x$,$y$) which appear on the RHS of the transformation rules. 
This formalism is much simpler than general query languages, such as SQL, and RDB2RDF mapping languages, such as R2RML~\cite{HRG11}, but it is expressive enough to specify object preserving views, the class of views we focus on in this article. 

\begin{table*}
\caption{Concrete predicates used by the TR Patterns in Table~\ref{t-ca}} \label{t-bp}
\begin{tabular*}{\textwidth}{l @{\extracolsep{\fill}}lll}
\hline\noalign{\smallskip}
\textbf{Built-in predicate}  							&&\textbf{Intuitive definition}\\
\noalign{\smallskip}\hline\noalign{\smallskip}
\textit{nonNull}($v$)								&&\textit{nonNull}($v$) holds iff value $v$ is not null\\[4pt]
\textit{RDFLiteral}($u$, \At, \Rn, $v$)					&&Given a value $u$, an attribute \At\ of \Rn, a relation name \Rn,\\
											&&and a literal $v$, \textit{RDFLiteral}($u$, \At, \Rn, $v$) holds iff $v$ is the\\
											&&literal representation of $u$, given the type of \At\ in \Rn\\[5pt]
\Fk(\Ti, $s$) 									&&Given a tuple \Ti\ of \Rn\ and a tuple $s$ of $S$, \Fk(\Ti, $s$) holds iff \Ti \\
where \Fk\ is a foreign key of the  					&& is related to $s$ by a foreign key \Fk \\ 
form \Fk(\Rn:$L$, $S$:$K$)						&& \\[5pt]
\textit{hasURI}(\Pt, \At, $s$) 						&&Given a tuple \Ti\ of \Rn, \textit{hasURI}(\Pt, \At, $s$) holds iff $s$ is the  \\
where \Pt\ is the namespace 						&&URI obtained  by concatenating the namespace prefix \Pt \\
prefix  and \At\ is a list of 							&&and  the attribute values $a_1, \dots, a_n$ where \At\ is the list  \\
attributes  of \Rn								&&(in Prolog notation) [$a_1, \dots, a_n$].  To further simply mat-  \\
											&& ters, we admit denoting a list with a  single element, ``[$a$]",  \\
											&&simply as ``$a$".\\
\noalign{\smallskip}\hline
\end{tabular*}
\end{table*}
Queries \Qn($x$) and \Qn($x$, $y$) are expressed as a list of literals. 
A literal can be: 
(1) a \textit{range expression} of the form \Rn(\Ti), 
where \Rn\ is a relation name in $U$ and \Ti\ is a tuple variable in \textit{\textbf{T}}; 
(2) a \textit{built-in predicate} or \textit{function}, 
such as those in Table~\ref{t-bp}.
The inclusion of built-in predicates and functions allows the formalism to capture specific notions of concrete domains, such as ``string concatenation'' and ``less than'', required for the specification of complex mappings and restrictions. 

\begin{table}
\caption{Transformation Rules} \label{t-TRs}
\begin{tabular}{lll}
\hline\noalign{\smallskip}
\textbf{TR}&& \textbf{Transformation Rules}\\
\noalign{\smallskip}\hline\noalign{\smallskip}
CTR	&& $\psi$: \Cn(x) $\leftarrow$ \Rn(\Ti), B[\Ti, x], where\\
        && - $\psi$ is the name of the CTR. \\
        && - \Cn\ is a class in $V$ and $x$ is a scalar variable whose value is a URI.\\
        && - \Rn\ is relation in \Sii\ and \Ti\ is tuple variable; \Rn\ is called the pivot relation and \\
        &&  \hspace{0.3cm}\Ti\ the pivot tuple variable of the rule.\\
        && - $B$[\Ti, $x$] is a list of literals.\\[5pt]
DTR	&& $\psi$: \Pt(x, y) $\leftarrow$ \Rn(\Ti), B[\Ti, x], H[\Ti, y], where\\
	&& - $\psi$ is the name of the DTR. \\
        && - \Pt\ is a datatype property in $V$ with domain \Do.\\
        && - ``R(\Ti), $B$[\Ti, $x$]'' is the right-hand side for the CTR that matches class $D$ \\
        && \hspace{0.3cm} with pivot relation \Rn. \\
        && - $H$[\Ti, $y$] is a list of literals which define a predicate $H$ that relates a tuple \Ti\  \\
        && \hspace{0.3cm}and data values in $y$.\\[5pt]
OTR		&& $\psi$: \Pt(x, y) $\leftarrow \Rn_\Do(\Ti_1), B_\Do[\Ti_1, x], H[\Ti_1, \Ti_2], \Rn_G(\Ti_2), B_G[\Ti_2, y]$, where\\
		&& - $\psi$ is the name of the OTR. \\
        && - \Pt\ is an object property in $V$ with domain \Do\ and range $G$.\\
        && - ``\Rn$_\Do$(\Ti$_1$), $B_\Do$[\Ti$_1$, $x$]'' is the RHS of the CTR $\psi_\Do$ that matches class $D$ with \\
        && \hspace{0.3cm} pivot relation \Rn$_\Do$. \\
        && - ``\Rn$_G$(\Ti$_2$), $B_G$[\Ti$_2$, $y$]'' is the RHS of the CTR $\psi_G$ that matches class $G$ with  \\
        && \hspace{0.3cm}pivot relation \Rn$_G$. \\
        && - $H$[\Ti$_1$, \Ti$_2$] is a list of literals which define a predicate $H$ that relates tuples in  \\
        && \hspace{0.3cm}\Rn$_\Do$ with tuples in \Rn$_G$.\\
\noalign{\smallskip}\hline
\end{tabular}
\end{table}
In this article, we adopt three specific types of transformation rules, 
defined in Table~\ref{t-TRs}: \textit{Class Transformation Rule} (CTR), \textit{Datatype Property Transformation Rule} (DTR), and \textit{Object Property Transformation Rule} (OTR).

In a transformation rule, a \emph{pivot relation} is the relation whose tuples
serve as the starting point (\emph{pivot tuples}) for the construction of RDF
triples. In particular, pivot tuples determine the URI of the subject of the
triples generated by the rule.

Intuitively, a CTR $\psi$ maps tuples of \Rn\ into 
instances of class \Cn\ in $V$. 
The predicate $B$[\Ti, $x$] establishes a 
\textit{semantic equivalence relation} between a tuple \Ti\ in \Rn, called the \textit{pivot tuple}, 
and an instance $x$ of \Cn\ 
(i.e., intuitively, \Ti\ and $x$ represent the same real-world entity). 

Since we are only interested in object-preserving views, 
the predicate $B$[\Ti,$x$] should define a partial one-to-one function: 
(a) each pivot tuple \Ti\ in \Rn\ should correspond to 
at most one instance $x$ of $C$; 
(b) different pivot tuples \Ti$_1$, \Ti$_2$ should correspond to 
different instances $x_1$, $x_2$ of $C$. 
We say that \Ti\ and $x$ are \textit{semantically equivalent}, 
denoted \Ti\ $\equiv$ $x$, w.r.t the CTR $\psi$.

Intuitively, a DTR $\psi$  defines values of a datatype property for instances generated from the
pivot relation by extracting attribute values either directly from the pivot tuple
or from tuples reachable through a relational path in the body of the rule. Therefore,  the values of 
the datatype property \Pt\  may correspond to attributes of the pivot tuple \Ti, or attributes of tuples related to \Ti, as specified by $H$ (see Table~\ref{t-TRs}).
Here, there is no restriction on the predicate $H$, which may associate several values $y$ to the same tuple \Ti.

An OTR defines instances of an object property by relating instances generated
from the pivot relation to instances reached through a relational path specified
in the body of the rule.
To interpret an OTR $\psi$, remember that the right-hand side of 
the CTRs $\psi_\Do$ and $\psi_G$ define instances of classes $D$ and $G$, respectively.
So, the OTR $\psi$ maps relations between tuples of \Rn$_\Do$ and \Rn$_G$ to instances of the object property \Pt, as specified by $H$.
Again, there is no restriction on the predicate $H$.

The approach proposed in this article efficiently computes changesets
by exploring the object-preserving property, 
which allows the precise identification of the tuples in 
the pivot relations whose corresponding instances 
may have been affected by an update. 
The advantages of this approach are: 
(1) it simplifies the specification of the view, 
as the formalism is specifically designed to describe 
the correspondences that define an object-preserving view; 
(2) the restrictions on the structure of the queries help ensure 
the consistency of the specification of the view; 
(3) the formal expressions that define the queries 
can be explored to construct the procedures automatically 
that compute the changesets that maintain the RDB2RDF view; 
(4) it facilitates the task of providing rigorous proofs 
for the correctness of the proposed approach. 

The problem of generating transformation rules is addressed in~\cite{VCC13,VCMN14} and is outside of the scope of this work. 
However, we summarize in Table~\ref{t-ca}  
a set of Transformation Rule (TR) patterns 
that lead to the definition of relational to RDF mappings
that guarantees that the RDF views satisfy the object-preserving property 
by construction. 
Table~\ref{t-bp} shows the definitions of the concrete predicates used by the TR Patterns in Table~\ref{t-ca}. 
The TR Patterns support most types of data restructuring commonly encountered when transforming relational data to RDF, and they suffice to capture all R2RML mapping patterns proposed in the literature~\cite{SPV12,DSC12}. 
In~\cite{VCMN14}, the authors proposed an approach to automatically generate R2RML mappings, based on a set of TR patterns. 
The approach uses relational views as a middle layer, which facilitates the R2RML generation process and improves the maintainability and consistency of the mapping. 

\begin{table}
\caption{Transformation Rule Patterns} \label{t-ca}
\begin{tabular}{lll}
\hline\noalign{\smallskip}
\textbf{TR}&& \textbf{Transformation Rule Pattern}\\
\noalign{\smallskip}\hline\noalign{\smallskip}
CTR		&& $\psi$:\Cn($s$) $\leftarrow$ \Rn(\Ti), \textit{hasURI}($P$, \At, $s$), \sel(\Ti), where\\
		&& - \Rn\ is a relation name in \Sii\ and \Ti\ a is tuple variable associated with \Rn,\\
		&& - \At\ is a list of attributes of a primary key of \Rn,\\
		&& - \sel\ is an optional selection over \Rn, and \\
		&& - $P$ is a namespace prefix	 \\[5pt]

OTR		&& $\psi$:\Pt($s$, $o$) $\leftarrow$ \Rn$_\Do$(\Ti), $B_\Do$[\Ti, $s$],\Fk$_1$(\Ti, \Ti$_1$), $\dots$, \Fk$_{n}$(\Ti$_{n-1}$, \Ti$_n$), \Rn$_G$(\Ti$_n$), $B_G$[\Ti$_n$, $o$], \\
		&& where:\\
		&& - \Pt\ is an object property of $V$,\\
		&& - \Rn\ is a relation name in \Sii\ and \Ti\ is a tuple variable associated with \Rn,\\
		&& - [\Fk$_1$, $\dots$, \Fk$_{n}$] is a path from \Rn\ to relation \Rn$_n$ where \Fk$_1$ relates \Rn\ and \Rn$_1$, and    \\
		&& \hspace{2mm}\Fk$_i$ relates \Rn$_{i-1}$ and \Rn$_i$, and \Ti$_{i}$ is a tuple variable associated with \Rn$_i$, 1$<$ i$\leq$n\\[5pt]
DTR		&&$\psi$:\Pt($s$, $v$) $\leftarrow$ \Rn(\Ti), $B$[\Ti, $s$], \Fk$_1$(\Ti, \Ti$_1$), $\dots$, \Fk$_{n}$(\Ti$_{n-1}$, \Ti$_n$), \textit{nonNull}(\Ti$_n$.\At$_1$), $\dots$,\\
		&& \hspace{17mm}\textit{nonNull}(\Ti$_n$.\At$_k$), \textit{RDFLiteral}(\Ti$_n$.\At$_1$, "\At$_1$", "\Rn$_n$", $v_1$), $\dots$, \\
		&& \hspace{17mm}\textit{RDFLiteral}(\Ti$_n$.\At$_k$, "\At$_k$", "\Rn$_n$", $v_k$), $T$([$v_1$, $\dots$, $v_k$], $v$), where\\
		&& - \Pt\ is a datatype property of $V$,\\
		&& - \Rn\ is a relation name in \Sii\ and \Ti\ a is tuple variable associated with \Rn,\\
		&& - [\Fk$_1$, $\dots$, \Fk$_{n}$] is a path from \Rn\ to relation \Rn$_n$ where \Fk$_1$ relates \Rn\ and \Rn$_1$, and    \\
		&& \hspace{2mm}\Fk$_i$ relates \Rn$_{i-1}$ and \Rn$_i$, and \Ti$_{i}$ is a tuple variable associated with \Rn$_i$, 1$<$ i$\leq$n\\
		&& - \At$_1$, $\dots$, \At$_k$ are the attributes of \Rn$_n$. If there is no path, then \At$_1$, $\dots$, \At$_k$ are  \\
		&& \hspace{2mm}attributes  of \Rn\ \\
		&& - $T$ is an optional function that transforms values of attributes \At$_1$, $\dots$, \At$_k$ to  \\
		&& \hspace{2mm}values of property \Pt \\
\noalign{\smallskip}\hline
\end{tabular}
\end{table}
\section{Case Study: \textbf{\textit{MusicBrainz\_RDF}}}\label{s-cs}

\begin{figure}
\center
\subfigure[]{\includegraphics[width=0.45\textwidth]{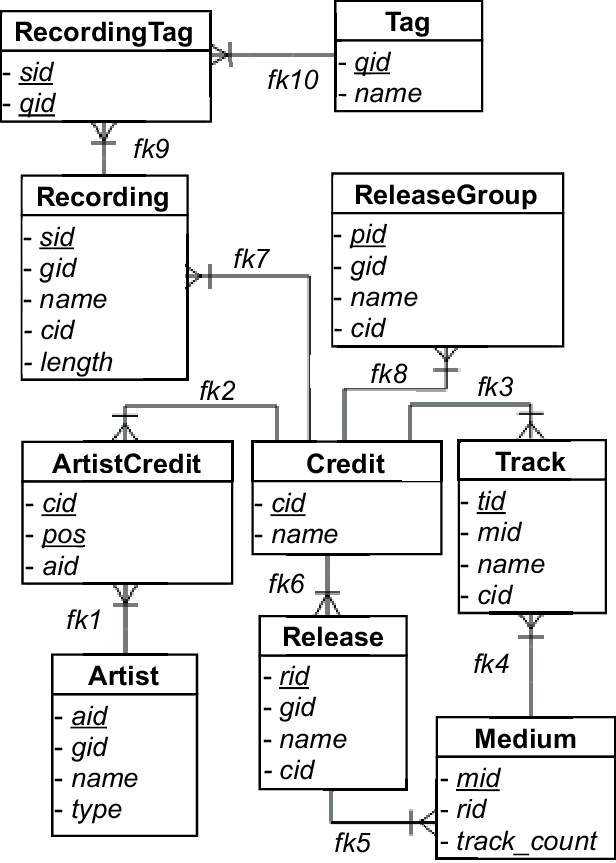}}
\qquad
\subfigure[]{\includegraphics[width=0.45\textwidth]{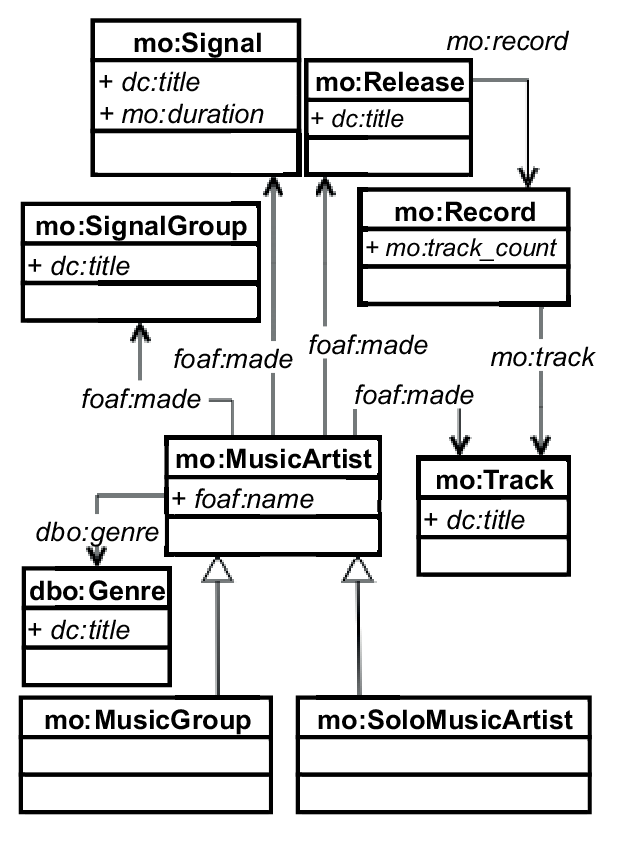}}
\caption{(a) Fragment of \textit{\textbf{MusicBrainz}} Schema and (b) Fragment of \textit{\textbf{MusicBrainz\_RDF}} View Ontology.}
\label{f-rdb}
\end{figure}

\textit{MusicBrainz}~\cite{MBz} is an open music encyclopedia that collects music metadata. 
The \textit{\textbf{MusicBrainz}} relational database is built on the PostgreSQL relational database engine and contains all of MusicBrainz music metadata. 
This data includes information about artists, release groups, releases, recordings, works, and labels, along with the many relationships among them. 
Figure~\ref{f-rdb}(a) depicts a fragment of the \textit{\textbf{MusicBrainz}} relational database schema (for more information about the original scheme, see~\cite{MBzS}). 
Each relation has a distinct primary key, whose name ends with ``id", except \textit{gid}, which is an \ac{UUID} for use in permanent links and external applications.
The relations \textit{Artist, Medium, Release, Recording} and \textit{Track}, in Figure~\ref{f-rdb}(a) represent the main concepts. 
The relation \textit{ArtistCredit} represents an N:M relationship between \textit{Artist} and \textit{Credit}. 
The labels of the arcs, such as \textit{fk1}, are the names of the foreign keys. 

\begin{figure}[h!]
        \centering
	\includegraphics[width=\textwidth]{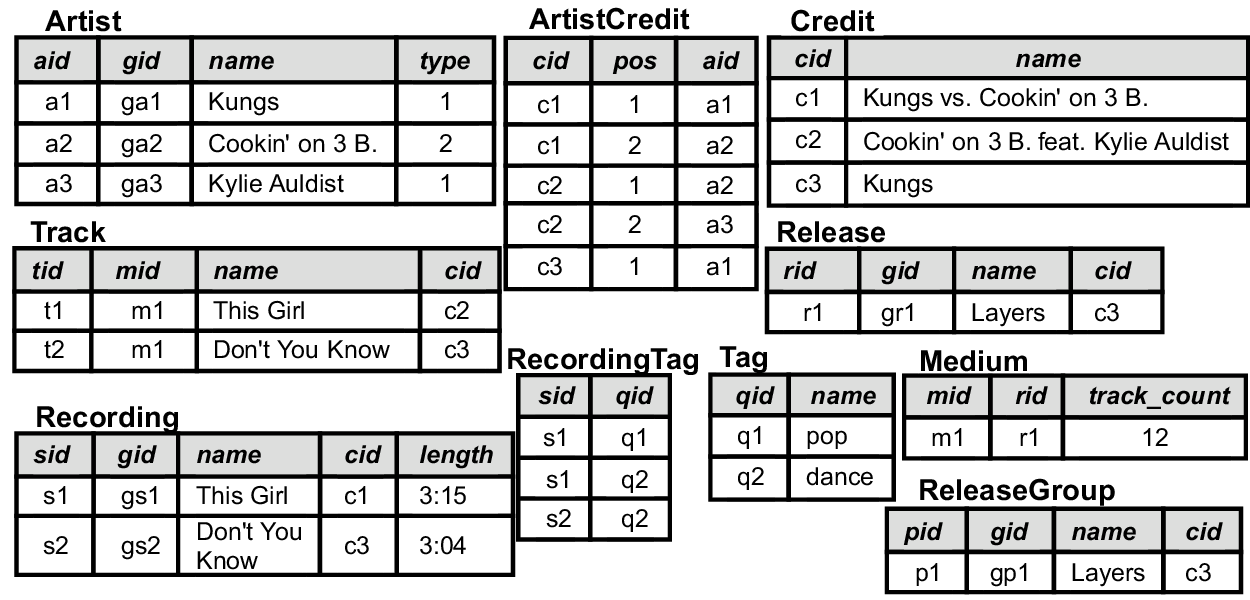}
	\caption{State Example for Relations in Figure~\ref{f-rdb}(a).}\label{f-rdbS}
\end{figure}

The case study uses an RDB2RDF view, 
called \textbf{\textit{MusicBrainz\_RDF}}, 
which is defined over the relational schema in Figure~\ref{f-rdb}(a). 
Appendix A provides the Data Definition Language (DDL) for the tables utilized in this case study. 

Figure~\ref{f-rdb}(b) depicts a fragment of the ontology used for publishing the \textbf{\textit{MusicBrainz\_RDF}} view. Appendix B shows the OWL ontology vocabulary used in the case study. 
It reuses terms from three well-known vocabularies, \textit{FOAF} (Friend of a Friend), \textit{MO} (Music Ontology) and \textit{DC} (Dublin Core).

Table~\ref{t-tr} shows a set of transformation rules that partially specify the mapping between the relational schema in Figure~\ref{f-rdb}(a) and the ontology in Figure~\ref{f-rdb}(b), obtained with the help of the tool described in~\cite{VCMN14}. The complete set of transformation rules for the specification of the 
\textbf{\textit{MusicBrainz\_RDF}}  view is defined in~\cite{Vidal20}
All RDF resources are generated using the auxiliary predicate
\textit{hasURI}, which constructs MusicBrainz-based URIs from the
primary keys of relational tuples, ensuring deterministic identifier
generation during the maintenance process.

For the examples in the following sections, consider the database state shown in Figure~\ref{f-rdbS}. 
The transformation rules $\psi_1$ and $\psi_2$,  in Table~\ref{t-tr}, are examples of the CTR Pattern. 
The predicate \textit{hasURI} is defined in Table~\ref{t-bp}.
The CTR $\psi_1$ indicates that, for each tuple \Ti\ in \textit{Artist}, one should: 
\begin{enumerate}
\item Compute the URI \uri\ such that \textit{hasURI}(\textit{mbz:}, \Ti.\textit{gid}, \uri)=true. 
\item Produce triple (\uri \hspace{0.5mm} \textit{rdf:type} \hspace{0.5mm} \textit{mo:MusicArtist}). 
Therefore, \Ti\ and \uri\ are semantically equivalent.
\end{enumerate}

\begin{table}
\caption{Transformation Rules} \label{t-tr}
\begin{tabular}{ll}
\hline\noalign{\smallskip}
\textbf{TR}&\textbf{Transformation Rules}\\
\noalign{\smallskip}\hline\noalign{\smallskip}
\rowcolor{gray!10}
\Ru$_1$	& \textit{mo:MusicArtist}(\uri) 		$\leftarrow$	\textit{Artist}(\Ti), \textit{hasURI}(\textit{mbz:},\Ti.\textit{gid}, \uri)\\[3pt]
\Ru$_2$	& \textit{mo:SoloMusicArtist}(\uri) 	$\leftarrow$	\textit{Artist}(\Ti), \textit{hasURI}(\textit{mbz:},\Ti.\textit{gid} , \uri), (\Ti.\textit{type} = 1)\\[3pt]
\rowcolor{gray!10}
\Ru$_3$	& \textit{mo:MusicGroup}(\uri) 		$\leftarrow$	\textit{Artist}(\Ti), \textit{hasURI}(\textit{mbz:}, \Ti.\textit{gid}, \uri), (\Ti.\textit{type} = 2)\\[3pt]
\Ru$_4$	& \textit{mo:Record}(\uri) 			$\leftarrow$	\textit{Medium}(\Ti), \textit{hasURI}(\textit{mbz:}, \Ti.\textit{mid}, \uri)\\[3pt]
\rowcolor{gray!10}
\Ru$_5$	& \textit{mo:Track}(\uri) 			$\leftarrow$	\textit{Track}(\Ti), \textit{hasURI}(\textit{mbz:}, \Ti.\textit{tid}, \uri)\\[3pt]
\Ru$_6$	& \textit{mo:Release}(\uri) 			$\leftarrow$	\textit{Release}(\Ti), \textit{hasURI}(\textit{mbz:}, \Ti.\textit{gid}, \uri)\\[3pt]
\rowcolor{gray!10}
\Ru$_7$&\textit{foaf:made}(\uri, $g$) 		$\leftarrow$	\textit{Artist}(\Ti), \textit{hasURI}(\textit{mbz:}, \Ti. \textit{gid}, \uri), \textit{fk1}(\Ti, \Ti1), \textit{fk2}(\Ti1, \Ti2),  \\
\rowcolor{gray!10}
		& \hspace{25mm}\textit{fk3}(\Ti2, \Ti3), \textit{Track}(\Ti3), \textit{\textit{hasURI}}(\textit{mbz:}, \Ti3.\textit{tid}, $g$) \\[3pt]
\Ru$_8$	& \textit{mo:track} (\uri, $g$) 		$\leftarrow$	\textit{Medium}(\Ti), 	\textit{hasURI}(\textit{mbz:}, \Ti.\textit{mid}, \uri), \textit{fk4}(\Ti, $f$), \textit{Track}($f$), \\
		& \hspace{25mm}\textit{hasURI}(\textit{mbz:}, $f$.\textit{tid}, $g$),  \textit{hasURI}(\textit{mbz:}, $f$.\textit{tid}, $g$)\\[3pt] 
\rowcolor{gray!10}
\Ru$_9$	& \textit{mo:record}(\uri, $g$) 		$\leftarrow$	\textit{Release}(\Ti), \textit{hasURI}(\textit{mbz:}, \Ti.\textit{gid}, \uri), 	\textit{fk5}(\Ti, $f$), \textit{Medium}($f$), \\
\rowcolor{gray!10}
		& \hspace{25mm}\textit{hasURI}(\textit{mbz:}, $f$.\textit{mid}, $g$)\\[3pt]
\Ru$_{10}$& \textit{foaf:name}(\uri, $v$) 		$\leftarrow$ \textit{Artist}(\Ti), \textit{hasURI}(\textit{mbz:}, \Ti.$gid$, \uri), \textit{nonNull}(r.\textit{name}), \\
		& \hspace{24mm}\textit{RDFLiteral}(r.\textit{name}, "\textit{name}", "\textit{Artist}", $v$)    \\[3pt]
\rowcolor{gray!10}
\Ru$_{11}$& \textit{mo:track\_count}(\uri, $v$)$\leftarrow$\textit{Medium}(\Ti),\textit{hasURI}(\textit{mbz:}, \Ti. \textit{mid}, \uri),\textit{nonNull}(\Ti.\textit{track\_count}),\\
\rowcolor{gray!10}
                  & \hspace{30mm}\textit{RDFLiteral}(\Ti.\textit{track\_count}, "\textit{track\_count}", "\textit{Medium}", $v$)   \\[3pt]    
\Ru$_{12}$& \textit{dc:title}(\uri, $v$) 		$\leftarrow$ \textit{Track}(\Ti), \textit{hasURI}(\textit{mbz:},  \Ti.\textit{tid}, \uri), \textit{nonNull}(\Ti.\textit{name}),\\
		& \hspace{22mm}\textit{RDFLiteral}(\Ti.\textit{name}, "\textit{title}", "\textit{Track}", $v$)\\[3pt]
		\rowcolor{gray!10}
\Ru$_{13}$& \textit{dc:title}(\uri, $v$) 		$\leftarrow$ \textit{Release}(\Ti), \textit{hasURI}(\textit{mbz:}, \Ti.$gid$, \uri), \textit{nonNull}(\Ti.\textit{name}),\\
\rowcolor{gray!10}
		& \hspace{22mm}\textit{RDFLiteral}(\Ti.\textit{name}, "\textit{title}", "\textit{Release}", $v$)\\[3pt]   
\Ru$_{14}$& \textit{mo:Signal}(\uri) 			$\leftarrow$ \textit{Recording}(\Ti), \textit{hasURI}(\textit{mbz:}, \Ti.$gid$, \uri)\\[3pt] 
\rowcolor{gray!10}
\Ru$_{15}$& \textit{mo:SignalGroup}(\uri) 		$\leftarrow$ \textit{ReleaseGroup}(\Ti), \textit{hasURI}(\textit{mbz:}, \Ti.$gid$, \uri)\\[3pt]   
\Ru$_{16}$& \textit{dbo:Genre}(\uri) 			$\leftarrow$ \textit{Tag}(\Ti), \textit{hasURI}(\textit{mbz:}, \Ti.$gid$, \uri)\\[3pt] 
\rowcolor{gray!10}
\Ru$_{17}$& \textit{dc:title}(\uri, $v$) 		$\leftarrow$ \textit{Recording}(\Ti), \textit{hasURI}(\textit{mbz:},  \Ti.\textit{gid}, \uri), \textit{nonNull}(\Ti.\textit{name}), \\
\rowcolor{gray!10}
		& \hspace{21mm}\textit{RDFLiteral}(\Ti.\textit{name}, "\textit{title}", "\textit{Recording}", $v$)\\[3pt]     
\Ru$_{18}$& \textit{mo:duration}(\uri, $v$) 	$\leftarrow$ \textit{Recording}(\Ti), \textit{hasURI}(\textit{mbz:},  \Ti.\textit{mid}, \uri), \textit{nonNull}(\Ti.\textit{length}), \\
		& \hspace{28mm}\textit{RDFLiteral}(\Ti.\textit{length}, "\textit{duration}", "\textit{Recording}", $v$)\\[3pt]     
		\rowcolor{gray!10}
\Ru$_{19}$& \textit{dc:title}(\uri, $v$) 		$\leftarrow$ \textit{ReleaseGroup}(\Ti), \textit{hasURI}(\textit{mbz:},  \Ti.\textit{gid}, \uri), \textit{nonNull}(\Ti.\textit{name}), \\
\rowcolor{gray!10}
		& \hspace{20mm} \textit{RDFLiteral}(\Ti.\textit{name}, "\textit{title}", "\textit{ReleaseGroup}", $v$)\\[3pt]  
\Ru$_{20}$& \textit{dc:title}(\uri, $v$) 		$\leftarrow$ \textit{Tag}(\Ti), \textit{hasURI}(\textit{mbz:},  \Ti.\textit{gid}, \uri), \textit{nonNull}(\Ti.\textit{name}), \\
		& \hspace{22mm}\textit{RDFLiteral}(\Ti.\textit{name}, "\textit{title}", "\textit{Tag}", $v$)\\[3pt]  
		\rowcolor{gray!10}
\Ru$_{21}$& \textit{foaf:made}(\uri, $g$) 		$\leftarrow$	\textit{Artist}(\Ti), \textit{hasURI}(\textit{mbz:}, \Ti. \textit{gid}, \uri), \textit{fk1}(\Ti, \Ti1), \\
\rowcolor{gray!10}
		& \hspace{25mm}\textit{fk2}(\Ti1, \Ti2), \textit{fk6}(\Ti2, \Ti3), \textit{Release}(\Ti3), \textit{\textit{hasURI}}(\textit{mbz:}, \Ti3.\textit{gid}, $g$) \\[3pt]
\Ru$_{22}$& \textit{foaf:made}(\uri, $g$) 		$\leftarrow$	\textit{Artist}(\Ti), \textit{hasURI}(\textit{mbz:}, \Ti. \textit{gid}, \uri), \textit{fk1}(\Ti, \Ti1), \textit{fk2}(\Ti1, \Ti2),\\
		& \hspace{25mm} \textit{fk8}(\Ti2, \Ti3), \textit{ReleaseGroup}(\Ti3), \textit{\textit{hasURI}}(\textit{mbz:}, \Ti3.\textit{gid}, $g$) \\[3pt]
		\rowcolor{gray!10}
\Ru$_{23}$& \textit{foaf:made}(\uri, $g$) 		$\leftarrow$	\textit{Artist}(\Ti), \textit{hasURI}(\textit{mbz:}, \Ti. \textit{gid}, \uri), \textit{fk1}(\Ti, \Ti1), \textit{fk2}(\Ti1, \Ti2), \\
\rowcolor{gray!10}
		& \hspace{24mm} \textit{fk7}(\Ti2, \Ti3), \textit{Recording}(\Ti3), \textit{\textit{hasURI}}(\textit{mbz:}, \Ti3.\textit{gid}, $g$) \\[3pt]
\Ru$_{24}$& \textit{dbo:genre}(\uri, $g$) 		$\leftarrow$	\textit{Artist}(\Ti), \textit{hasURI}(\textit{mbz:}, \Ti. \textit{gid}, \uri), \textit{fk1}(\Ti, \Ti1), \textit{fk2}(\Ti1, \Ti2),  \\
		& \hspace{25mm}\textit{fk7}(\Ti2, \Ti3), \textit{fk9}(\Ti3, \Ti4), \textit{fk10}(\Ti4, \Ti5),\textit{Tag}(\Ti5),  \\
		& \hspace{25mm}\textit{\textit{hasURI}}(\textit{mbz:}, \Ti5.\textit{qid}, $g$)\\
\noalign{\smallskip}\hline
\end{tabular}
\end{table}

The CTR $\psi_2$ indicates that, for each tuple \Ti\ in \textit{Artist},
where \Ti.type = 1, one should: 
\begin{enumerate}
\item Compute the URI \uri\ such that \textit{hasURI}(\textit{mbz:},\Ti.\textit{gid}, \uri)=true. 
\item Produce triple (\uri\ \hspace{0.5mm} \textit{rdf:type} \hspace{0,5mm} \textit{mo:SoloMusicArtist}). 
Therefore, \Ti\ and \uri\ are semantically equivalent.
\end{enumerate} 

Note that attribute \textit{gid} is a key for relation \textit{Artist}. 
Therefore, different tuples in \textit{Artist} generate distinct URIs, i.e., distinct instances. 
Also, note that $\psi_1$ and $\psi_2$ use the same predicate \textit{hasURI}. 
Therefore, if a tuple \Ti\ is mapped to triples ($x$ \hspace{0.5mm} \textit{rdf:type} \hspace{0.5mm} \textit{mo:MusicArtist}) and ($y$ \hspace{0.5mm} \textit{rdf:type} \hspace{0.5mm} \textit{mo:SoloMusicArtist}), then $x$=$y$.

Considering the database state in Figure~\ref{f-rdbS}, CTRs $\psi_1$ and $\psi_2$ produce the following triples:  

\noindent(\textit{mbz:}\Ti.\textit{ga1} \hspace{2mm}\textit{rdf:type} \hspace{2mm}\textit{mo:MusicArtist});
(\textit{mbz:}\Ti.\textit{ga1} \hspace{1mm}\textit{rdf:type} \hspace{1mm}\textit{mo:SoloMusicArtist});
(\textit{mbz:}\Ti.\textit{ga2} \hspace{2mm}\textit{rdf:type} \hspace{2mm}\textit{mo:MusicArtist});
(\textit{mbz:}\Ti.\textit{ga3} \hspace{2mm}\textit{rdf:type} \hspace{2mm}\textit{mo:MusicArtist}); and (\textit{mbz:}\Ti.\textit{ga3} \hspace{2mm}\textit{rdf:type} \hspace{2mm}\textit{mo:SoloMusicArtist}).

The transformation rule $\psi_{10}$, in Table~\ref{t-tr}, is an example of the DTR Pattern. 
The predicates \textit{nonNull}($v$) and \textit{RDFLiteral}($u$, \At, \Rn, \vl) are defined in Table~\ref{t-bp}.
Rule $\psi_{10}$ matches the value of attribute \textit{name} of relation \textit{Artist} with the value of datatype property \textit{foaf:name}, whose domain is \textit{mo:MusicArtist}.
It indicates that, for each tuple \Ti\ of \Rn, one should: 
\begin{enumerate}
\item Compute the URI \uri\ for the instance of \textit{mo:Music\-Artist} that \Ti\ represents, using the CTR $\psi_{1}$. Therefore, \uri\ and \Ti\ are semantically equivalent.
\item For each value \vl, where \vl\ is the literal representation of \Ti.\textit{name} and \Ti.\textit{name} is not NULL, produce triple (\uri\ \hspace{0.5mm} \textit{foaf:name} \hspace{0.5mm}  \vl).
\end{enumerate}

Considering the database state in Figure~\ref{f-rdbS}, $\psi_{10}$ produces the following triples: 

\noindent(\textit{mbz:ga1} \hspace{2mm}\textit{foaf:name}  \hspace{2mm} ``Kungs'');
(\textit{mbz:ga2}  \hspace{2mm}\textit{foaf:name}  \hspace{2mm} ``Cookin's on 3 B.''); and
(\textit{mbz:ga3}  \hspace{2mm}\textit{foaf:name}  \hspace{2mm} ``Kylie Auldist'').

The transformations rules $\psi_7$ and $\psi_8$, in Table~\ref{t-tr}, are examples of the OTR Pattern.
The predicate \Fk(\Ti, \uri), where \Fk\ is a foreign key of the form \Fk(\Rn:$L$, $S$:$K$), is defined in Table~\ref{t-bp}. 
For example, rule $\psi_7$ matches a relationship between a tuple \Ti\ of \textit{Artist} and a tuple \Ti$_3$ in \textit{Track} to instances of the object property \textit{foaf:made}, whose domain is \textit{mo:MusicArtist} and range is \textit{mo:Track}.
It indicates that, for each tuple \Ti\ of \Rn, one should: 
\begin{enumerate}
\item Compute the URI \uri\ for the instance of \textit{mo:Music\-Artist}  that \Ti\ represents, using the CTR  $\psi_1$. 
Therefore, \uri\ and \Ti\ are semantically equivalent.
\item For each tuple \Ti$_3$ of \textit{Track} such that \Ti\ is related to \Ti$_3$  through path [$fk_1, fk_2, fk_3$], compute the URI $g$ for the instance of \textit{mo:Track} that \Ti$_3$ represents (using CTR $\psi_5$).
Therefore, \Ti$_3$ and $g$ are semantically equivalent.
\item Produce triple (\uri\ \hspace{0.5mm} \textit{foaf:made} \hspace{0.5mm} $g$)
\end{enumerate}

Considering the database state in Figure~\ref{f-rdbS}, $\psi_7$  produces the following triples:

\noindent(\textit{mbz:ga1}\hspace{2mm} \textit{foaf:made}\hspace{2mm} \textit{mbz:t2});
(\textit{mbz:ga2}\hspace{2mm} \textit{foaf:made}\hspace{2mm} \textit{mbz:t1}); and \\
(\textit{mbz:ga3}\hspace{2mm} \textit{foaf:made}\hspace{2mm} \textit{mbz:t1}).
\section{Materialization of an RDB2RDF view} \label{s-mgr}

The materialization of the data graph for an RDB2RDF view requires translating source data into the RDB2RDF view vocabulary as specified by the mappings. 
An important technical issue that arises in this process is the possibility of \textit{duplicated triples}, i.e., triples generated more than once due to different assignments to variables in the body of one or more transformation rules. 
Indeed, the main difficulty in incrementally maintaining views with duplicates lies in delete and update operations. 
Recall that, by definition, the same triple cannot be present twice in an RDF triple store. 
Thus, if a tuple is removed, we cannot determine whether the corresponding triples should be deleted from the view, because triples may still be produced by another tuple in the database. 

For a proper handling of the issue of duplicates, we distinguish between two types of duplication: (1) duplicated triples generated from different pivot relations (see Table~\ref{t-TRs} for the definition of pivot relation); (2) duplicated triples generated from the same pivot relation. 

For the case of duplicated triples generated from different pivot relations, we propose a solution based on named graphs. In the proposed framework, the content of an RDF2RDB view is stored in an RDF dataset that contains a collection of named graphs. 
As in Carroll et al.~\cite{Carroll05}, a \textit{named graph} is defined as a pair comprising a URI and an RDF graph.  
A named graph can be considered a set of quadruples (or ``quads'') with the subject, predicate, and object of the triples as the first three components, and the graph URI as the fourth element. 
Each quadruple is interpreted similarly to a triple in RDF, except that the predicate denotes a ternary relation, instead of a binary relation. 
This way of representing quadruples, called \textit{quad-statements}, was incorporated in the specification of \textit{N-Quad~}~\cite{nQuads}.

The main reason for separating triples into distinct (named) graphs is that duplicated triples, produced by tuples in different relations, will be in different named graphs (context). 
This is an important property for supporting duplicated triples generated by different tuples.

For the case of duplicated triples generated by the same pivot relation, our solution requires detecting the pivot tuples that may be affected by an update, and then re-materializing all triples produced by those tuples.  
In particular, we may characterize the approach as that of ``tracking the relevant tuples in the pivot relations for a given update" rather than ``tracking the updated triples in the view for a given update". 

The following definitions formalize the materialization of the view 
\Vi = ($V$\!, \Sii, \Mv) as the result of applying 
the \acp{TR} in \Mv\ against a state $\sigma$ of database \Sii.

For the examples in this section, consider the RDB2RDF view \linebreak \textit{\textbf{Mu\-sic\-Brainz\_RDF}} defined in Section~\ref{s-cs}, and the database state shown in Figure~\ref{f-rdbS}.
Also, consider that mbz:ga, mbz:gm, mbz:gr, and mbz:gt are the named graph URIs for the pivot relations \textit{Artist, Medium} and \textit{Track}, respectively.

In the rest of the article, $R(\sigma)$ denotes the relation associated
with the relation scheme $R$ in the database state $\sigma$. 

\begin{definition}\label{d-5.1}
Let \Ru\ be a \acs{TR} in \Mv\ and \Ti$_1, \dots, $\Ti$_n$ be tuple variables appearing in \Ru\ associated with relations \Rn$_1, \dots,$ \Rn$_n$, where $n$ $\geq$ 2 and \Ti$_i$ $\neq$ \Ti$_j$ for i $\neq$ j.
Also, let $\sigma$ be a database state and let $p_1, \dots, p_n$ be tuples in \Rn$_1$($\sigma$), $\dots$, \Rn$_n$($\sigma$). 

\begin{enumerate}
\item  \Ru[\Ti$_1$/$p_1$ , $\dots$, \Ti$_n$/$p_n$] denotes the \ac{TR} obtained from \Ru\ 
by substituting the tuples $p_1$, $\dots$, $p_n$  for the tuple variables \Ti$_1, \dots,$ \Ti$_n$, respectively.
\item  \Ru[\Ti$_1$/$p_1$ , $\dots$, \Ti$_n$/$p_n$]($\sigma$) denotes the set of triples which are produced  when \Ru[\Ti$_1$/$p_1$, $\dots$, \Ti$_n$/$p_n$] is applied to $\sigma$. 
\end{enumerate}
\end{definition}

\begin{definition}[RDF State of a Pivot Tuple under a Rule]
Let $\Psi$ be a transformation rule whose pivot relation is $R^*$, and let $\sigma$ be a relational database state. 
For any tuple $p \in R^*$, the function
\[
RDF\_State[\Psi](p, \sigma)
\]
returns the set of RDF quads generated by applying the transformation rule $\Psi$ to the pivot tuple $p$, evaluated over the database state $\sigma$.

Formally,
\[
RDF\_State[\Psi](p, \sigma) = 
\]
 \{(\uri, $q$, $o$, $g$) $|$ (\uri, $q$, $o$) is a triple in \Ru[\Ti/$p$]($\sigma$) and $g$ is the named graph URI for pivot relation \Rn\}.
 
\end{definition}

\begin{definition}[RDF State of a Pivot Tuple]
Let $\mathcal{M}$ be the set of transformation rules defining an RDF view,
and let $R^*$ be a pivot relation of some rule in $\mathcal{M}$.
Let $\sigma$ be a relational database state.

For any tuple $p \in R^*(\sigma)$, the RDF state of $p$ under state $\sigma$ is defined as

\[
RDF\_State[R^*](p, \sigma)
=
\bigcup_{\Psi \in \mathcal{M},\, pivot(\Psi)=R^*}
RDF\_State[\Psi](p, \sigma).
\]
\end{definition}

Note that, if $p$ is a tuple in \Rn($\sigma$) such that \Rn\ is not a pivot relation in any \ac{TR} in \Mv,
then \Mv[$p$]($\sigma$) = $\emptyset$.


\begin{example} 
\label{e-5.1}
Consider the transformation rules for the \textbf{\textit{MusicBrainz\_RDF}} view defined in Table~\ref{t-tr}. 
Also, consider the relation \textit{Artist} in Figure~\ref{f-rdb}(a), which is the pivot relation of the \acp{TR}  \Ru$_1$, \Ru$_2$, \Ru$_3$, \Ru$_7$, \Ru$_{10}$,  and \Ru$_{24}$.
Thus, the RDF state of a tuple $p$ in relation \textit{Artist} is computed by applying the \acp{TR} \Ru$_1$[\Ti/$p$],  \Ru$_2$[\Ti/$p$],  \Ru$_3$[\Ti/$p$],  \Ru$_7$[\Ti/$p$], \Ru$_{10}$[\Ti/$p$], and \Ru$_{24}$[\Ti/$p$]. 
Considering the database state in Figure~\ref{f-rdbS}, the RDF state of tuple \textit{a1} in \textit{Artist} contains the following quads:
\begin{itemize}
\item (mbz:ga1 \hspace{1mm}\textit{rdf:type} \hspace{1mm}\textit{mo:MusicArtist} \hspace{1mm} mbz:ga), by \Ru$_1$[\Ti/$a1$]
\item (mbz:ga1 \hspace{1mm}\textit{rdf:type} 	\hspace{1mm}\textit{mo:SoloMusicArtist} \hspace{1mm} mbz:ga), by \Ru$_2$[\Ti/$a1$]
\item (mbz:ga1 \hspace{2.5mm}\textit{foaf:made} \hspace{2mm}mbz:t2 \hspace{2mm} mbz:ga), by \Ru$_7$[\Ti/$a1$]
\item (mbz:ga1 \hspace{1mm}\textit{foaf:name} \hspace{1mm}"Kungs" \hspace{1mm} mbz:ga), by \Ru$_{10}$[\Ti/$a1$]
\item (mbz:ga1 \hspace{2.5mm}\textit{dbo:genre} \hspace{2mm}mbz:q1 \hspace{2mm} mbz:ga); \\
(mbz:ga1 \hspace{2.5mm}\textit{dbo:genre} \hspace{2mm}mbz:q2 \hspace{2mm} mbz:ga);\\
(mbz:ga1 \hspace{2.5mm}\textit{dbo:genre} \hspace{2mm}mbz:q2 \hspace{2mm} mbz:ga), by \Ru$_{24}$[\Ti/$a1$]
\end{itemize}
  
Note that, \Ru$_3$[\Ti/$a1$] produces no triple, while \Ru$_{24}$[\Ti/$a1$] produces duplicated triples. 
The triple (mbz:ga1 \hspace{2.5mm}\textit{dbo:genre} \hspace{2mm}mbz:gq2) is generated twice by assigning different tuple variables in \Ru$_{24}$.
\end{example}

\begin{definition}[RDF State of an RDB2RDF View]
Let $W$ be an RDB2RDF view defined by a set of transformation rules $\mathcal{M}$, 
and let $\sigma$ be a relational database state. 
Let $\mathcal{P}(W)$ denote the set of pivot relations of $W$.

The RDF state of $W$ under state $\sigma$, denoted by $RDF\_State(W,\sigma)$, 
is defined as the union of the RDF states of all pivot tuples of all pivot relations of $W$:

\[
RDF\_State(W,\sigma)
=
\bigcup_{R^* \in \mathcal{P}(W)}
\;\;
\bigcup_{p \in R^*(\sigma)}
RDF\_State[R^*](p,\sigma).
\]
\end{definition}

Intuitively, the RDF state of an RDB2RDF view of \Vi\ at state $\sigma$ is computed by the materialization of  RDF states of all pivot relations of \Vi. 


%
%
\section{Formal Framework for Computing Correct Changesets for RDB2RDF Views} \label{s-fw}

This Section presents the proposed framework for computing a correct changeset for the materialized RDB2RDF view \Vi = ($V$\!, \Sii, \Mv), when an update u occurs in the source relational database \Sii.

\subsection{Overview}

An update \up\ on a relation \Rn\ is defined as two sets, $D$ and $I$, of tuples of \Rn. 
The update \up\ indicates that the tuples in $D$ must be deleted and the tuples in $I$ must be inserted into \Rn. 
More precisely, we have:

\begin{definition}[updates, insertions and deletions]\label{d-up}
\mbox{}
An \textit{update} on a relation \Rn\ is a pair \up=($D, I$) such that
$D$ and $I$ are, possibly empty, sets of tuples of \Rn.
If $I$ = $\emptyset$, we say that the update is a \textit{deletion}, and, if $D$=$\emptyset$, we say that the update is an insertion.
\end{definition}

The semantics of an update \up\ = ($D$, $I$) is straightforward and is defined as follows:

\begin{definition}[semantics of an update]\label{d-induc}
\mbox{}
Let \up\ = ($D$, $I$) be an update on \Rn\ and \St$_0$ be a database state. 
Then, the \textit{execution} of \up\ on $\St_0$ results in the database state $\St_1$ which is equal to \St$_0$ except that 
\Rn(\St$_1$) = \Rn(\St$_0$)$ - D \cup I$. 
We also say that \St$_1$ \textit{is the result of executing} \up\ on \St$_0$ 
and that \St$_0$ is the database state \textit{before} \up\ and \St$_1$ is the database state \textit{after} \up.
\end{definition}

Note that updates are deterministic in the sense that, given an initial state, an update always results in the same state.

The proposed approach to compute the correct changeset for an update \up\ follows three main steps: 

\begin{enumerate}
\item \textbf{Identification of \textit{Relevant Relations}.} 
Identify the relations in \Sii\ that are relevant to update \up. 
A relation is considered to be relevant to \up\ if its RDF state is possibly affected by $u$.  

\item \textbf{Identification of \textit{Relevant Tuples}.}
\label{l-irt} 
Identify the tuples in the relevant relations that are relevant to the update.
A tuple is considered to be relevant to an update \up\ if its RDF state is possibly affected by the update.

\item \textbf{Computation of Changesets.} 
Compute the changeset $\langle$\Dm(\up), \Dp(\up)$\rangle$. 
\Dm(\up) contains the old  RDF states of the relevant tuples, which are removed from \Vi, and \Dp(\up) contains the new RDF states of the relevant tuples, which are inserted into \Vi.
Therefore, only the RDF state of the relevant tuples identified in Step~\ref{l-irt} is rematerialized.
\end{enumerate}

\begin{algorithm}
\begin{minipage}{\hsize}
\footnotesize{
\hspace{-0.5cm}

\Indentp{-1em}
	\KwIn{\\
	\Indentp{1.4em}
	\up\ = ($D$, $I$) $-$ an update on \Rn; \\
	\St$_0$ and \St$_1$ $-$ the states of the database respectively before and after the update \up;}
	\KwOut{\\
	\Indentp{1.4em}
	\Dm\ and \Dp } 
\Indentp{1.4em}
	
\hspace{-0.5cm}
{
\hspace{-0.5cm}

\textbf{Phase 1:} Before the update do:  \\

\hspace{0.5cm}
1.1 Compute \Pc$_0$, the set of tuples in $\St_0$\ that are relevant to \up\ before the update  (Definition~\ref{def:rtb})  \\

\hspace{0.5cm}
1.2 Compute \Dm := $\underset{p \in (\Pc_0)} \bigcup
RDF\_State(p, \St_0$\ );

\hspace{0.5cm}\tcp{\Dm\ contains the union of the RDF states of tuples in \Pc$_0$ (Def.~\ref{def:delta-minus})}

\hspace{-0.5cm}

\textbf{Phase 2:} After the update do:  \\
	
\hspace{0.5cm}
2.1 Compute \Pc$_1$,  the set of tuples in $\St_1$\ that are relevant to \up\ (Def.~\ref{def:rtb})  \\

\hspace{0.5cm}
2.2 Compute \Dp := $\underset{p \in (\Pc_1)} \bigcup 
RDF\_State(p, \St_1$\ );
		
\hspace{0.5cm}\tcp{\Dp\ contains the union of the RDF states of tuples in \Pc$_1$ (Def.~\ref{def:delta-minus})}
}
\Return(\Dp,\Dm);
}

\end{minipage}%
		\caption{Algorithm for computing changeset for updates on a relation \Rn}\label{f-upC_alg}
\end{algorithm}

Algorithm~\ref{f-upC_alg} shows a high-level description 
of the algorithm for computing changeset for updates on a relation \Rn, where \Rn~is relevant to the view W.  
The process consists of two phases. 
Phase 1 uses the database state before the update, while Phase 2 uses the database state after the update. 
The algorithms for insertions and deletions are defined similarly and omitted here.

\subsection{ Formal Definitions for Computing Changesets}

In the following, we present the precise definitions of the key concepts for computing changesets.

\maketitle

\begin{definition}[Transformation Rules Relevant to a Relation]
Let $M$ be a set of transformation rules and let $R$ be a relation scheme.
A transformation rule $\Psi \in M$ with pivot relation $R^{*}$ is relevant to $R$ iff $R \in Relations$ ($path(\Psi)$. 
We denote by $RelevantTRs(M, R)$ the set of transformation rules in $M$
that are relevant to $R$.
\end{definition}


\begin{definition}[Impacted Transformation Rules]
Let $M$ be a set of transformation rules, and let $R$ and $R^*$ 
be relation schemes. The set of transformation rules of pivot $R^*$ 
that are impacted by updates on relation $R$, denoted by
\[
ImpactedTRs(R^*, R),
\]
is defined as
\[
ImpactedTRs(R^*, R)
=
\{\Psi \in M 
\mid 
pivot(\Psi) = R^* 
\;\land\;
R \in Relations(path(\Psi))
\}.
\]
\end{definition}


\begin{definition}[Relevant Tuples Before Update (RTB)]
\label{def:rtb}
Let $u=(D,I)$ be an update on a relation schema $R$, and let $\sigma_0$ be the
database state before applying $u$.
Let $M$ be the set of transformation rules.

The set of relevant tuples before the update, denoted by $RTB(u)$, is the set of
pairs $(r^{*},R^{*})$ such that $r^{*}\in R^{*}(\sigma_0)$ and there exists a
transformation rule $\Psi \in RelevantTRs(M,R)$ with pivot relation $R^{*}=pivot(\Psi)$
satisfying one of the following conditions:

\begin{enumerate}
\item[(i)] \textbf{Relevance via a relevant path ($R^{*}\neq R$):}
There exists a prefix path $\phi$ of $path(\Psi)$ from $R^{*}$ to $R$ and a tuple
$t \in D$ such that
\[
r^{*} \in P[\phi](t;\sigma_0).
\]

\item[(ii)] \textbf{Pivot-relation case ($R^{*}=R$):}
$R^{*}=R$ and
\[
r^{*}\in D.
\]
\end{enumerate}
\end{definition}


\begin{definition}[Changeset $\Delta^{-}$ for an update]
\label{def:delta-minus}

Let $u=(D,I)$ be an update on a relation $R$, let $\sigma_0$ denote the database state before applying $u$, and let $M$ be the set of transformation rules.

Let $\mathrm{RTB}(u)$ be the set of pairs $(p^{\ast},R^{\ast})$ relevant to $u$ before the update, as defined in Definition~\ref{def:rtb}.

Let $\mathrm{RDF\_State}[R](p;\sigma)$ be the function that computes the RDF state of a tuple $p$ of relation $R$ evaluated over a database state $\sigma$.

The \emph{changeset before the update}, denoted $\Delta^{-}(u)$, is defined as:
\[
\Delta^{-}(u)
\;=\;
\bigcup_{(p^{\ast},R^{\ast}) \in \mathrm{RTB}(u)}
\mathrm{RDF\_State}[R^{\ast}](p^{\ast};\sigma_0).
\]
\end{definition}


\begin{definition}[Relevant Tuples After Update (RTA)]
\label{def:rtb_after}
Let $u=(D,I)$ be an update on a relation schema $R$, and let $\sigma_1$ be the
database state AFTER applying $u$.
Let $M$ be the set of transformation rules.

The set of relevant tuples AFTER the update, denoted by $RTA(u)$, is the set of
pairs $(r^{*},R^{*})$ such that $r^{*}\in R^{*}(\sigma_1)$ and there exists a
transformation rule $\Psi \in RelevantTRs(M,R)$ with pivot relation $R^{*}=pivot(\Psi)$
satisfying one of the following conditions:

\begin{enumerate}
\item[(i)] \textbf{Relevance via a relevant path ($R^{*}\neq R$):}
There exists a prefix path $\phi$ of $path(\Psi)$ from $R^{*}$ to $R$ and a tuple
$t \in I$ such that
\[
r^{*} \in P[\phi](t;\sigma_1).
\]

\item[(ii)] \textbf{Pivot-relation case ($R^{*}=R$):}
$R^{*}=R$ and
\[
r^{*}\in I.
\]
\end{enumerate}
\end{definition}


\begin{definition}[Changeset $\Delta^{+}$ for an update]
\label{def:delta-plus}

Let $u=(D,I)$ be an update on a relation $R$, 

let $\sigma_1$ denote the database state before applying $u$, and 

let $M$ be the set of transformation rules.

Let $\mathrm{RTA}(u)$ be the set of pairs $(p^{\ast},R^{\ast})$ relevant to $u$ AFTER the update, as defined in Definition~\ref{def:rtb}.

Let $\mathrm{RDF\_State}[R](p;\sigma)$ be the function that computes the RDF state of a tuple $p$ of relation $R$ evaluated over a database state $\sigma$.

The \emph{changeset before the update}, denoted $\Delta^{+}(u)$, is defined as:
\[
\Delta^{+}(u)
\;=\;
\bigcup_{(p^{\ast},R^{\ast}) \in \mathrm{RTA}(u)}
\mathrm{RDF\_State}[R^{\ast}](p^{\ast};\sigma_1).
\]
\end{definition}

%
%
\subsection{Computation of the Changeset $\langle \Delta^{-}(u), \Delta^{+}(u) \rangle$}

Let $u = (D,I)$ be an update on relation $R$, producing a transition 
from database state $\sigma_0$ to $\sigma_1$.

In the following, based on the formal definitions introduced previously, we present the steps for computing $\Delta^{-}(u)$ and  $\Delta^{+}(u)$.

\subsubsection{Computation of $\Delta^{-}(u)$}

The computation of $\Delta^{-}(u)$ is performed in three steps:

\begin{enumerate}

\item \textbf{Identification of Relevant Transformation Rules.} \\
Determine the set of transformation rules $\Psi \in M$ that are relevant 
to updates on relation $R$, according to the formal definition of relevant TR. 
A rule $\Psi$ is relevant if and only if:
\begin{itemize}
    \item[(i)] $pivot(\Psi) = R$, or
    \item[(ii)] $R \in Relations(path(\Psi))$.
\end{itemize}

\item \textbf{Derivation of $RTB(u)$.} \\
Compute the set $RTB(u)$ (Relevant Tuples Before) as defined in Definition~XX.
For each relevant transformation rule $\Psi$:
\begin{itemize}
    \item If $pivot(\Psi) = R$, the affected pivot tuples correspond to 
    the tuples in the deletion component $D$.
    \item If $R \in Relations(path(\Psi))$, the affected pivot tuples are 
    obtained by evaluating the relational path of $\Psi$ over the 
    pre-update state $\sigma_0$.
\end{itemize}

\item \textbf{Computation of the Deletion Component.} \\
Finally, compute:
\[
\Delta^{-}(u)
=
\bigcup_{(p^*, R^*) \in RTB(u)}
RDF\_STATE[R^*](p^*; \sigma_0)
\]
That is, $\Delta^{-}(u)$ is defined as the union of the RDF states of all pivot tuples in $RTB(u)$, evaluated over the pre-update database state $\sigma_0$.

\end{enumerate}

\subsubsection{Computation of $\Delta^{+}(u)$}

The computation of $\Delta^{+}(u)$ follows a symmetric structure, but is evaluated entirely over the post-update state $\sigma_1$:

\begin{enumerate}

\item \textbf{Identification of Relevant Transformation Rules.} \\
Determine the same set of relevant transformation rules $\Psi \in M$ 
for updates on relation $R$.

\item \textbf{Derivation of $RTA(u)$.} \\
Compute the set $RTA(u)$ (Relevant Tuples After).
For each relevant transformation rule $\Psi$:
\begin{itemize}
    \item If $pivot(\Psi) = R$, the affected pivot tuples correspond to 
    the tuples in the insertion component $I$.
    \item If $R \in Relations(path(\Psi))$, the affected pivot tuples are 
    obtained by evaluating the relational path of $\Psi$ over the 
    post-update state $\sigma_1$.
\end{itemize}

\item \textbf{Computation of the Insertion Component.} \\
Compute:
\[
\Delta^{+}(u)
=
\bigcup_{(p^*, R^*) \in RTA(u)}
RDF\_STATE[R^*](p^*; \sigma_1)
\]
Thus, $\Delta^{+}(u)$ is the union of the RDF states of all pivot tuples in $RTA(u)$, evaluated over the post-update database state $\sigma_1$.

\end{enumerate}

%
%
\subsection{Case Study: Computing Changeset for \textit{\textbf{MusicBrainz\_RDF}} view }

\subsubsection{Identifying Relevant Transformation Rules}

\begin{example}\label{e-5.3}
Consider the transformation rules for the \textit{\textbf{MusicBrainz\_RDF}} view defined in Table~\ref{t-tr}. Also consider an update on relation \textit{Track}.The relation \textit{Track} is relevant to \acp{TR} \Ru$_5$, \Ru$_7$ and \Ru$_8$ (Definition~\ref{def:rtb}). 
The relations \textit{Track, Artist} and \textit{Medium} are the pivot relations of \Ru$_5$, \Ru$_7$, and \Ru$_8$, respectively. 
Therefore, the relations \textit{Track, Artist} and \textit{Medium} are relevant to updates on the relation \textit{Track} (Definition~\ref{def:rtb}).
\end{example}

\subsubsection{Identifying Relevant Tuples}

\
\begin{example}\label{e-5.4}
Consider the transformation rules for the \textit{\textbf{MusicBrainz\_RDF}} view defined in Table~\ref{t-tr}, 
and the database state in Figure~\ref{f-rdbS}. 
Also, consider \up\ an update which deletes tuple \Ti$_{old}$ and inserts tuple \Ti$_{new}$ in table \textit{Track}, where:
\begin{itemize}
\item \Ti$_{old}$ = $\langle$t1, m1, ``This Girl", c2$\rangle$
\item \Ti$_{new}$=$\langle$t1, m1, ``This Girl (feat. Cookin' On 3 B.)", c1$\rangle$ 
\end{itemize}

From \acp{TR} \Ru$_7$ , \Ru$_8$, we have that the relations \textit{Track, Artist} and \textit{Medium} are relevant to updates on table \textit{Track} (See Example~\ref{e-5.3}).

\noindent From Definition~\ref{def:rtb} and TR \Ru$_7$, we have that:
\begin{itemize}
\item \Pc[\textit{Track}, \Ru$_7$](\Ti$_{new}$)= \{\textit{a1}, \textit{a2}\}
\item \Pc[\textit{Track}, \Ru$_7$](\Ti$_{old}$)= \{\textit{a2}, \textit{a3}\}.
\end{itemize}
Therefore, tuples \textit{a1} and \textit{a2} in relation \textit{Artist} are related to \Ti$_{new}$ w.r.t. \Ru$_7$, and tuples \textit{a2} and \textit{a3} in relation \textit{Artist} are related to \Ti$_{old}$ w.r.t. \Ru$_7$. Thus,  tuples \textit{a1, a2} and \textit{a3} are relevant to update \up\ w.r.t. \Ru$_7$. 

From Definition~\ref{def:rtb} and TR \Ru$_8$, we have that:
\begin{itemize}
\item \Pc[\textit{Track}, \Ru$_8$](\Ti$_{new}$)= \{\textit{m1}\}
\item \Pc[\textit{Track}, \Ru$_8$](\Ti$_{old}$)= \{\textit{m1}\}.
\end{itemize}
Therefore, tuple \textit{m1} in relation \textit{Medium} is relevant to update \up\ w.r.t \Ru$_8$.
From Definition~\ref{def:rtb}(i), tuples \textit{a1, a2, a3} and  \textit{m1} are relevant to \up. Since table \textit{Track} is a pivot relation, from Definition~\ref{def:rtb}(ii), \Ti$_{new}$ and \Ti$_{old}$ are also relevant to \up.
\end{example}

\subsubsection{Computing Changesets}

\label{s-remat}

\begin{example}\label{e-5.5}
To illustrate this strategy, consider the update \up\ as in Example~\ref{e-5.3}, and \Pc$_0$ and \Pc$_1$ are in Definition~\ref{def:delta-minus}.
Figure~\ref{f-rdbNS} shows the new state of database \Sii\ after the update \up.
\begin{figure}[tb]
\centering
\includegraphics[width=\textwidth]{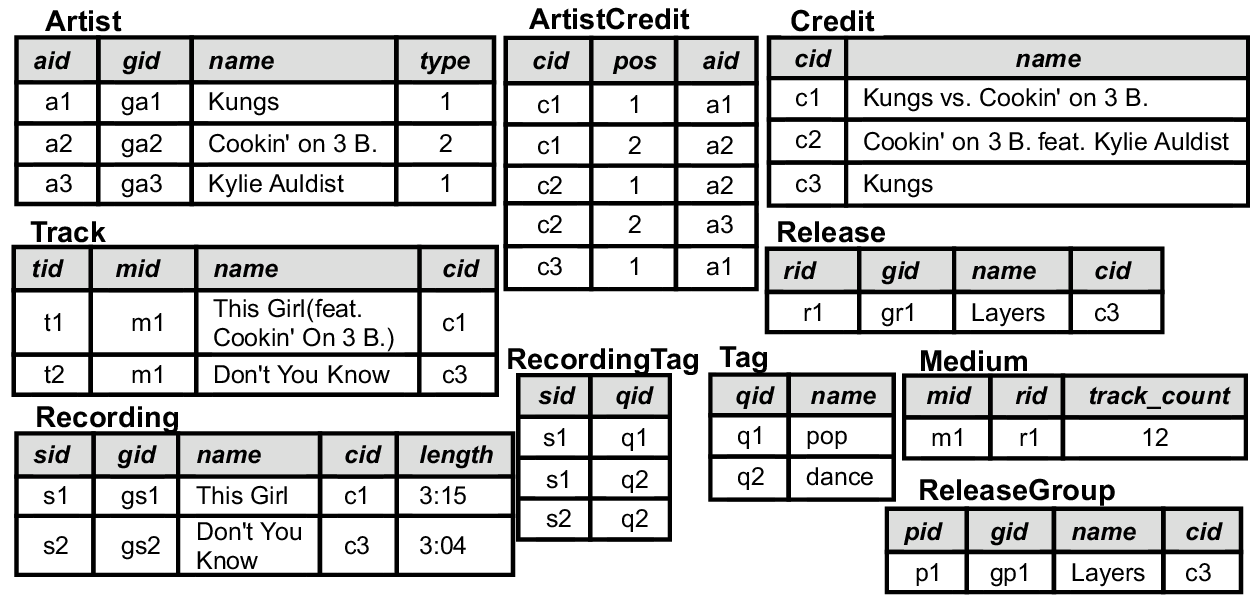}
\caption{Database State after the update \up.}\label{f-rdbNS}
\end{figure}
From Example~\ref{e-5.4},  we have:
\begin{eqnarray}
\Pc_0 = \{\mathit{a1}, \mathit{a2}, \mathit{a3},\mathit{m1}, \Ti_\text{old}\} \text{ and }
\Pc_1 = \{\mathit{a1}, \mathit{a2}, \mathit{a3},\mathit{m1}, \Ti_\text{new}\}.
\end{eqnarray}

Figure~\ref{f-Dm1} shows the set \Dm(\up), which contains the old RDF states of the tuples in \Pc$_0$, while Figure~\ref{f-Dp1} shows the set \Dp(\up), which includes the new RDF states of the tuples in \Pc$_1$.

\subsection{Generating Triggers to Compute Changesets}

In the proposed strategy, database triggers compute and publish the correct changeset for the RDB2RDF view, ensuring that the RDF materialization remains synchronized with the underlying relational database.

The first step of the strategy is to identify the \emph{relations in the source database that are relevant to the RDB2RDF view}, that is, those relations whose updates may affect the RDF state of the view (see Definition~XX).

For each update operation $u = (D, I)$ performed on a relevant relation $R$, a single statement-level \texttt{AFTER} trigger is defined.

The \texttt{AFTER} trigger is fired immediately after the execution of the update and is responsible for computing:

\begin{itemize}
    \item $\Delta^{-}(u)$: the set of RDF quads to be removed from the old RDF state $\sigma_0$, and
    \item $\Delta^{+}(u)$: the set of RDF quads to be inserted into the new RDF state $\sigma_1$.
\end{itemize}

Although $\Delta^{-}(u)$ is formally defined over the pre-update database state $\sigma_0$, its computation is performed inside an \texttt{AFTER} trigger, where the database is already in the post-update state $\sigma_1$. Correctness is ensured by explicitly enforcing $\sigma_0$-semantics during the evaluation.

In the pivot-relation case, where $\mathrm{pivot}(\Psi)=R$, the required information from $\sigma_0$ is directly available through the set $D$ of deleted tuples, which contains the old versions of the updated tuples of $R$.

In the relational-path case, where $R \in \mathrm{Relations}(\mathrm{path}(\Psi))$ and $\mathrm{pivot}(\Psi) \neq R$, the body of the rule must be evaluated under $\sigma_0$. Since relation $R$ is already in state $\sigma_1$ when the trigger executes, its pre-update state must be reconstructed explicitly. Let $I$ denote the set of inserted tuples. Then the old state of $R$ is obtained as:

\[
\sigma_0(R) = \big( R(\sigma_1) \setminus I \big) \cup D.
\]

Thus, even though the trigger executes after the update, the combination of the deleted and inserted tuple sets with the current database state allows the exact reconstruction of $\sigma_0(R)$ when required. Consequently, $\Delta^{-}(u)$ is computed precisely as defined over the pre-update database state.

Figure~\ref{fig:generic_after_trigger} presents the generic statement-level \texttt{AFTER} trigger used to compute the changeset $\langle \Delta^{-}(u), \Delta^{+}(u) \rangle$ for an update on relation $R$.

\begin{figure}[h]
\centering
\begin{verbatim}
CREATE TRIGGER t_delta_R
AFTER INSERT OR UPDATE OR DELETE ON R
REFERENCING OLD TABLE AS deleted_R
            NEW TABLE AS inserted_R
FOR EACH STATEMENT
EXECUTE FUNCTION trg_compute_delta_R_stmt();

CREATE OR REPLACE FUNCTION trg_compute_delta_R_stmt()
RETURNS trigger
LANGUAGE plpgsql
AS $$
BEGIN
  -- Phase 1 (conceptual sigma0): compute Delta-(u)
  PERFORM trg_compute_delta_minus_R_stmt_internal();

  -- Phase 2 (sigma1): compute Delta+(u)
  PERFORM trg_compute_delta_plus_R_stmt_internal();

  RETURN NULL;
END;
$$;
\end{verbatim}
\caption{Generic \texttt{AFTER} statement-level trigger to compute the changeset $\langle \Delta^{-}(u), \Delta^{+}(u) \rangle$ for an update $u = (D, I)$ on relation $R$.}
\label{fig:generic_after_trigger}
\end{figure}

\begin{figure}
\centering
\fbox{
\footnotesize{
\begin{tabular}{l}
\{(mbz:t1 \hspace{2mm}	\textit{rdf:type} \hspace{2mm} \textit{mo:track} \hspace{2mm} mbz:gt);\\
(mbz:t1 	\hspace{2mm}	\textit{dc:title} \hspace{2mm} ``This Girl'' \hspace{2mm} mbz:gt); \\
(mbz:ga2 	\hspace{2mm}	\textit{rdf:type} \hspace{2mm} \textit{mo:MusicArtist} \hspace{2mm} mbz:ga);\\
(mbz:ga2 	\hspace{2mm}	\textit{foaf:name} \hspace{2mm}``Cookin's on 3 B.'' \hspace{2mm} mbz:ga);\\ 
(mbz:ga2 	\hspace{2mm}	\textit{foaf:made} 	\hspace{2mm} mbz:t1 \hspace{2mm} mbz:ga);\\
(mbz:ga2 	\hspace{2mm}	\textit{dbo:genre}	\hspace{2mm}mbz:q1 \hspace{2mm} mbz:ga);\\
(mbz:ga2 	\hspace{2mm}	\textit{dbo:genre}	\hspace{2mm}mbz:q2 \hspace{2mm} mbz:ga);\\
(mbz:ga2 	\hspace{2mm}	\textit{rdf:type} 		\hspace{2mm} \textit{mo:MusicGroup} \hspace{2mm} mbz:ga);\\
(mbz:ga3 	\hspace{2mm}	\textit{rdf:type} 		\hspace{2mm} \textit{mo:MusicArtist} \hspace{2mm} mbz:ga); \\
(mbz:ga3 	\hspace{2mm}	\textit{foaf:name} 	\hspace{2mm}``Kylie Auldist'' \hspace{2mm} mbz:ga); \\
(mbz:ga3 	\hspace{2mm}	\textit{foaf:made} 	\hspace{2mm} mbz:t1 \hspace{2mm} mbz:ga); \\
(mbz:ga3 	\hspace{2mm}	\textit{rdf:type} 		\hspace{2mm} \textit{mo:SoloMusicArtist} \hspace{2mm} mbz:ga);\\ 
(mbz:m1	\hspace{2mm} 	\textit{rdf:type} 		\hspace{2mm} \textit{mo:Record} \hspace{2mm} mbz:gm);\\
(mbz:m1	\hspace{2mm}	\textit{mo:track\_count}\hspace{2mm} 12 \hspace{2mm} mbz:gm); \\
(mbz:m1 	\hspace{2mm}	\textit{mo:track} 		\hspace{2mm} mbz:t1 \hspace{2mm} mbz:gm);\\ 
(mbz:m1 	\hspace{2mm}	\textit{mo:track} 		\hspace{2mm} mbz:t2 \hspace{2mm} mbz:gm);\\
(mbz:ga1	\hspace{2mm}	\textit{rdf:type} 		\hspace{2mm} \textit{mo:MusicArtist} \hspace{2mm} mbz:ga);\\ 
(mbz:ga1 	\hspace{2mm}	\textit{foaf:name} 	\hspace{2mm}``Kungs'' \hspace{2mm} mbz:ga); \\
(mbz:ga1 	\hspace{2mm}	\textit{foaf:made} 	\hspace{2mm}mbz:t2 \hspace{2mm} mbz:ga) ; \\
(mbz:ga1 	\hspace{2mm}	\textit{dbo:genre} 	\hspace{2mm}mbz:q1 \hspace{2mm} mbz:ga);\\
(mbz:ga1 	\hspace{2mm}	\textit{dbo:genre} 	\hspace{2mm}mbz:q2 \hspace{2mm} mbz:ga);\\
(mbz:ga1 	\hspace{2mm}	\textit{rdf:type} 		\hspace{2mm}\textit{mo:SoloMusicArtist} \hspace{2mm} mbz:ga)
\}
\end{tabular}
}
}
\caption{\Dm(\up) for update \up\ in example~\ref{e-5.5}.}
\label{f-Dm1} 
\end{figure}

\begin{figure}
\centering
\fbox{
\footnotesize{
\begin{tabular}{l}
(mbz:t1 	\hspace{2mm}	\textit{rdf:type} 		\hspace{2mm}\textit{mo:track} \hspace{2mm} mbz:gt); \\
(mbz:t1 	\hspace{0.5mm}	\textit{dc:title} 	\hspace{0.5mm}``ThisGirl(feat. Cookin'On 3B.)" \hspace{0.5mm}mbz:gt);\\
\{(mbz:ga2	\hspace{2mm} 	\textit{rdf:type} 	\hspace{2mm} \textit{mo:MusicArtist} \hspace{2mm} mbz:ga); \\
(mbz:ga2 	\hspace{2mm}	\textit{foaf:name} 	\hspace{2mm}``Cookin's on 3 B." \hspace{2mm} mbz:ga);\\ 
(mbz:ga2	\hspace{2mm} 	\textit{foaf:made} 	\hspace{2mm}mbz:t1 \hspace{2mm} mbz:ga); \\
(mbz:ga2 	\hspace{2mm}	\textit{dbo:genre}	\hspace{2mm}mbz:q1 \hspace{2mm} mbz:ga);\\
(mbz:ga2 	\hspace{2mm}	\textit{dbo:genre}	\hspace{2mm}mbz:q2 \hspace{2mm} mbz:ga);\\
(mbz:ga2 	\hspace{2mm}	\textit{rdf:type} 		\hspace{2mm}\textit{mo:MusicGroup} \hspace{2mm} mbz:ga);\\
(mbz:ga3 	\hspace{2mm}	\textit{rdf:type} 		\hspace{2mm}\textit{mo:MusicArtist} \hspace{2mm} mbz:ga); \\
(mbz:ga3 	\hspace{2mm}	\textit{foaf:name} 	\hspace{2mm}``Kylie Auldist" \hspace{2mm} mbz:ga); \\
(mbz:ga3 	\hspace{2mm}	\textit{rdf:type} 		\hspace{2mm}\textit{mo:SoloMusicArtist} \hspace{2mm} mbz:ga);\\ 
(mbz:m1 	\hspace{2mm}	\textit{rdf:type} 		\hspace{2mm}\textit{mo:Record} \hspace{2mm} mbz:gm); \\
(mbz:m1 	\hspace{2mm}	\textit{mo:track\_count}\hspace{2mm} 12 \hspace{2mm} mbz:gm); \\
(mbz:m1 	\hspace{2mm}	\textit{mo:track} 	\hspace{2mm}mbz:t1 \hspace{2mm} mbz:gm);\\ 
(mbz:m1 	\hspace{2mm}	\textit{mo:track} 	\hspace{2mm}mbz:t2 \hspace{2mm} mbz:gm);\\
(mbz:ga1 	\hspace{2mm}	\textit{rdf:type} 		\hspace{2mm}\textit{mo:MusicArtist} \hspace{2mm} mbz:ga);\\ 
(mbz:ga1 	\hspace{2mm}	\textit{foaf:name} 	\hspace{2mm}``Kungs" \hspace{2mm} mbz:ga);\\
(mbz:ga1 	\hspace{2mm}	\textit{foaf:made} 	\hspace{2mm}mbz:t1 \hspace{2mm} mbz:ga); \\
(mbz:ga1 	\hspace{2mm}	\textit{foaf:made} 	\hspace{2mm}mbz:t2 \hspace{2mm} mbz:ga) ;\\
(mbz:ga1 	\hspace{2mm}	\textit{dbo:genre}	\hspace{2mm}mbz:q1 \hspace{2mm} mbz:ga);\\
(mbz:ga1 	\hspace{2mm}	\textit{dbo:genre}	\hspace{2mm}mbz:q2 \hspace{2mm} mbz:ga);\\
(mbz:ga1 	\hspace{2mm}	\textit{rdf:type} 		\hspace{2mm}\textit{mo:SoloMusicArtist} \hspace{2mm} mbz:ga)
\}
\end{tabular}
}
}
\caption{\Dp(\up) for update \up\ in example~\ref{e-5.5}.}
\label{f-Dp1} 
\end{figure}

%
\end{example}
\newpage
\section{Conclusions and Final Remarks }
\label{s-conc}
This article presented a formal framework for computing correct changesets for RDB2RDF views. 
In the proposed formal framework, changesets are computed in three steps: identification of relevant transformation rules, identification of relevant tuples, and Computation of Changesets.
Furthermore, the changesets are computed based solely on the update and the source database state, that is, the view is self-maintainable. 

The formal framework was based on three key ideas. 
First, it assumed that the RDB2RDF views are object-preserving, that is, they preserve the base entities of the source database rather than creating new entities from existing ones~\cite{Motschnig00}.
This assumption enables precise identification of the tuples relevant to a data source update w.r.t. an RDB2RDF view. 
Second, the formal framework included a rule language to specify object-preserving view mappings. Third, the proposed framework assumes that the content of an RDB2RD view is stored in an RDF dataset containing a set of named graphs that describe the context in which the triples were produced. 
The central result of the article showed that changesets computed according to the formal framework correctly maintain the RDB2RDF views.

Finally, we are currently developing a tool to automatically generate triggers for computing correct changesets for the RDB2RDF view, based on the view mappings. 

\begin{acknowledgements}
This work was partly funded by FAPERJ, grant E-26/204.322/2024, and 
by CNPq grant 305587/2021-8.
\end{acknowledgements}

%
%

\bibliographystyle{spmpsci}      
\bibliography{references}   

\appendix
\input{apendice}
%

%

\end{document}

%% file: apendice.tex

\section*{Appendix A -- Relational Schema}\label{ap-a}
\label{appendix:ddl}

\begin{lstlisting}[caption={Relational schema (DDL)},label={lst:musicbrainz-schema}]
CREATE TABLE Artist (
  aid TEXT PRIMARY KEY,
  gid UUID NOT NULL,
  name VARCHAR NOT NULL,
  type INTEGER
);

CREATE TABLE Credit (
  cid TEXT PRIMARY KEY,
  name VARCHAR NOT NULL
);

CREATE TABLE ArtistCredit (
  cid TEXT NOT NULL,
  pos INTEGER NOT NULL,
  aid TEXT NOT NULL,
  PRIMARY KEY (cid, pos),

  CONSTRAINT fk2 FOREIGN KEY (cid) REFERENCES Credit(cid),
  CONSTRAINT fk1 FOREIGN KEY (aid) REFERENCES Artist(aid)
);

CREATE TABLE Release (
  rid TEXT PRIMARY KEY,
  gid UUID NOT NULL,
  name VARCHAR NOT NULL,
  cid TEXT NOT NULL,

  CONSTRAINT fk6 FOREIGN KEY (cid) REFERENCES Credit(cid)
);

CREATE TABLE Medium (
  mid TEXT PRIMARY KEY,
  rid TEXT NOT NULL,
  track_count INTEGER NOT NULL,

  CONSTRAINT fk5 FOREIGN KEY (rid) REFERENCES Release(rid)
);

CREATE TABLE Track (
  tid TEXT PRIMARY KEY,
  mid TEXT NOT NULL,
  name VARCHAR NOT NULL,
  cid TEXT NOT NULL,

  CONSTRAINT fk4 FOREIGN KEY (mid) REFERENCES Medium(mid),
  CONSTRAINT fk3 FOREIGN KEY (cid) REFERENCES Credit(cid)
);

CREATE TABLE Recording (
  sid TEXT PRIMARY KEY,
  gid UUID NOT NULL,
  name VARCHAR NOT NULL,
  cid TEXT NOT NULL,
  length INTEGER,

  CONSTRAINT fk7 FOREIGN KEY (cid) REFERENCES Credit(cid)
);

CREATE TABLE Tag (
  qid TEXT PRIMARY KEY,
  name VARCHAR NOT NULL
);

CREATE TABLE RecordingTag (
  sid TEXT NOT NULL,
  qid TEXT NOT NULL,
  PRIMARY KEY (sid, qid),

  CONSTRAINT fk9 FOREIGN KEY (sid) REFERENCES Recording(sid),
  CONSTRAINT fk10 FOREIGN KEY (qid) REFERENCES Tag(qid)
);

CREATE TABLE ReleaseGroup (
  pid TEXT PRIMARY KEY,
  gid UUID NOT NULL,
  name VARCHAR NOT NULL,
  cid TEXT NOT NULL,

  CONSTRAINT fk8 FOREIGN KEY (cid) REFERENCES Credit(cid)
);
\end{lstlisting}

\section*{Appendix B -- Ontology Vocabulary (Classes and Properties)}\label{ap-b}
\begin{lstlisting}[style=appendixcode,language=ttl,caption={Ontology vocabulary (Turtle)},label={lst:ontology-ttl}]
@prefix mo:    <http://purl.org/ontology/mo/> .
@prefix foaf:  <http://xmlns.com/foaf/0.1/> .
@prefix dc:    <http://purl.org/dc/elements/1.1/> .
@prefix dbo:   <http://dbpedia.org/ontology/> .
@prefix muto:  <http://purl.org/muto/core#> .
@prefix mbz:   <http://musicbrainz.org/> .
@prefix rdfs:  <http://www.w3.org/2000/01/rdf-schema#> .
@prefix owl:   <http://www.w3.org/2002/07/owl#> .
@prefix xsd:   <http://www.w3.org/2001/XMLSchema#> .

##################################################
# Classes
##################################################

mo:MusicArtist a owl:Class ;
    rdfs:label "Music Artist" .

mo:SoloMusicArtist a owl:Class ;
    rdfs:subClassOf mo:MusicArtist ;
    rdfs:label "Solo Music Artist" .

mo:MusicGroup a owl:Class ;
    rdfs:subClassOf mo:MusicArtist ;
    rdfs:label "Music Group" .

mo:Track a owl:Class ;
    rdfs:label "Track" .

mo:Record a owl:Class ;
    rdfs:label "Record" .

mo:Release a owl:Class ;
    rdfs:label "Release" .

mo:Signal a owl:Class ;
    rdfs:label "Signal" .

mo:SignalGroup a owl:Class ;
    rdfs:label "Signal Group" .

dbo:Genre a owl:Class ;
    rdfs:label "Genre" .

muto:Tag a owl:Class ;
    rdfs:label "Tag" .

##################################################
# Object Properties
##################################################

foaf:made a owl:ObjectProperty ;
    rdfs:domain mo:MusicArtist ;
    rdfs:range owl:Thing ;
    rdfs:label "made" .

mo:track a owl:ObjectProperty ;
    rdfs:domain mo:Record ;
    rdfs:range mo:Track ;
    rdfs:label "track" .

mo:record a owl:ObjectProperty ;
    rdfs:domain mo:Release ;
    rdfs:range mo:Record ;
    rdfs:label "record" .

dbo:genre a owl:ObjectProperty ;
    rdfs:domain mo:MusicArtist ;
    rdfs:range dbo:Genre ;
    rdfs:label "genre" .

##################################################
# Datatype Properties
##################################################

foaf:name a owl:DatatypeProperty ;
    rdfs:domain owl:Thing ;
    rdfs:range xsd:string ;
    rdfs:label "name" .

dc:title a owl:DatatypeProperty ;
    rdfs:domain owl:Thing ;
    rdfs:range xsd:string ;
    rdfs:label "title" .

mo:duration a owl:DatatypeProperty ;
    rdfs:domain mo:Signal ;
    rdfs:range xsd:integer ;
    rdfs:label "duration" .

mo:track_count a owl:DatatypeProperty ;
    rdfs:domain mo:Record ;
    rdfs:range xsd:integer ;
    rdfs:label "track count" .
\end{lstlisting}